\documentclass[twocolumn]{article}

\usepackage[dvips]{epsfig}
\usepackage[]{amsbsy}

\def\thebibliography#1{\section*{\normalsize \bf References 
 }\list
 {[\arabic{enumi}]}{\settowidth\labelwidth{[#1]}\leftmargin\labelwidth
 \advance\leftmargin\labelsep
 \usecounter{enumi}}
 \def\newblock{\hskip .11em plus .33em minus .07em}
 \sloppy\clubpenalty4000\widowpenalty4000
 \sfcode`\.=1000\relax}

\begin{document}

\twocolumn[

\begin{center} \large \bf
   Dynamical mean-field study of the Mott transition in thin films
\end{center}
\vspace{-3mm}

\begin{center} 
   M. Potthoff and W. Nolting
\end{center}
\vspace{-6mm}

\begin{center} \small \it 
   Institut f\"ur Physik, 
   Humboldt-Universit\"at zu Berlin, 
   Germany
\end{center}
\vspace{2mm}

\begin{center}
\parbox{141mm}{ \small
The correlation-driven transition from a paramagnetic metal to
a paramagnetic Mott-Hubbard insulator is studied within
the half-filled Hubbard model for a thin-film geometry. We consider
simple-cubic films with different low-index surfaces and film
thickness $d$ ranging from $d=1$ (two-dimensional) up to $d=8$.
Using the dynamical mean-field theory, the lattice (film) problem
is self-consistently mapped onto a set of $d$ single-impurity 
Anderson models which are indirectly coupled via the respective 
baths of conduction electrons. The impurity models are solved at 
zero temperature using the exact-diagonalization algorithm. We 
investigate the layer and thickness dependence of the electronic 
structure in the low-energy regime. Effects due to the finite film 
thickness are found to be the more
pronounced the lower is the film-surface coordination number. 
For the comparatively open sc(111) geometry we find a strong
layer dependence of the quasi-particle weight
while it is much less pronounced for the sc(110) and the sc(100)
film geometries. For a given geometry and 
thickness $d$ there is a unique critical interaction strength 
$U_{c2}(d)$ at which all effective masses diverge and there is 
a unique strength $U_{c1}(d)$ where the insulating solution 
disappears. $U_{c2}(d)$ and $U_{c1}(d)$ gradually increase
with increasing thickness eventually approaching their bulk values.
A simple analytical argument explains the complete
geometry and thickness dependence of $U_{c2}$. $U_{c1}$ is found 
to scale linearly with $U_{c2}$.
}
\end{center}
\vspace{8mm} 
]



{\center \bf \noindent I. INTRODUCTION \\ \mbox{} \\} 

Electron-correlation effects in systems with thin-film geometry 
have gained increasing interest in condensed-matter physics. In 
particular, there has been intense research on thermodynamic phase 
transitions to a symmetry-broken (e.~g.\ magnetic) state below 
a critical temperature \cite{All94,Bab96}. In magnetic thin films 
the thickness dependence of the order parameter, of the critical 
temperature as well as of the critical exponents has been 
investigated both, experimentally \cite{DTP+89,SBS+90,LB92,FBS+93} 
and theoretically \cite{Hon90,EM91,JDB92,SSR96}.

A thin film in three dimensions belongs to a two-dimensional
universality class, regardless of the film thickness $d$ \cite{Fis73}.
Due to the Mermin-Wagner theorem \cite{MW66}, however, an effectively 
two-dimensional spin-isotropic system cannot display long-range 
magnetic order at any finite temperature. This is one important
reason why anisotropies play a fundamental role for the 
understanding of thermodynamic phase transitions in thin films.
The necessary inclusion of anisotropies, however, makes a theoretical
description considerably more complicated.

For a quantum phase transition the situation is different:
Symmetry breaking need not occur at the transition point,
and the energy scale that is characteristic for the transition 
at zero temperature, remains meaningful at any finite temperature.
Consequently, anisotropies are not vital for the understanding of
a quantum phase transition in thin films: The transition can 
be studied within an isotropic model and at any temperature,
starting from the monolayer ($d=1$) up to the three-dimensional 
limit ($d\mapsto \infty$).
From this point of view, following up the characteristics of a
quantum phase transition as a function of $d$, may be the better 
defined and the simpler problem if compared with a thermodynamic
transition.

One of the prime examples for a quantum phase transition 
is the correlation-driven transition from a paramagnetic metal 
to a paramagnetic insulator \cite{Mot61,Geb97}. 
Generally, the Mott transition is of interest since strong
electron correlations lead to low-energy electronic properties
that cannot be understood within an independent-electron picture;
the important correlation effects must be treated non-perturbatively.
Conventional band theory is unable to provide a satisfactory 
description of the transition.

While magnetic phase transitions in thin films are traditionally 
studied within localized-moment models (Ising or Heisenberg model)
\cite{Hon90,EM91,JDB92,SSR96},
the possibly simplest generic model for the Mott transition is 
the Hubbard model \cite{hubbard}. Contrary to the
localized-moment models, the Hubbard model describes itinerant 
electrons on a lattice which may form local moments as a consequence
of the strong on-site Coulomb interaction. Starting from the early 
approaches of Mott \cite{Mot61}, Hubbard \cite{Hub64b}, and Brinkman 
and Rice \cite{BR70}, there has been extensive work on the Mott 
transition in the Hubbard model (for a recent overview see Ref.\ 
\cite{Geb97}). A thin-film geometry has not been considered up to now.

The following particular questions shall be addressed in the present
study of the thin-film Mott transition:
First, the breakdown of translational symmetry in the film normal
direction introduces a layer dependence of physical quantities that 
characterize the transition. The layer dependence of the effective 
mass $m^\ast$ or the double occupancy $\langle n_\uparrow n_\downarrow
\rangle$, for example, is worth studying. 
Second, it has to be expected 
that there is a dependence on the film geometry. We can 
distinguish between surface effects and effects due to the
finite film thickness: Surface effects are already present
in thick films ($d \mapsto \infty$) and will be more 
pronounced for films with comparatively ``open'' 
surfaces, i.~e.\ for surfaces where the nearest-neighbor
coordination number is strongly reduced. For thin films
there may be finite-thickness effects in addition. Here, the
perturbation of the electronic structure introduced by 
one of the film 
surfaces affects the electronic structure in the vicinity of
the other one; both surfaces can ``interact'' with each other.
Third, it is not clear from 
the beginning whether or not there is a unique critical interaction 
strength $U_c$ at which the whole film undergoes the transition. 
Analogously to magnetic transitions where an enhanced surface 
critical temperature is discussed \cite{Mil71,MLS89,PN97b}, a modified 
$U_c$ for the film surface might exist for thicker films.
Finally, it is interesting to see the critical interaction
strength $U_c$ evolving as a function of $d$ and to investigate
how it approaches the bulk value. Again, the crossover from 
$D=2$ to $D=3$ dimensions should depend on the particular film 
geometry.

In the recent years, comprehensive investigations of the Mott
transition have been performed \cite{Geb97,GKKR96} for the 
Hubbard model in infinite spatial dimensions \cite{MV89,Vol93}. 
The $D=\infty$ model is amenable to an exact solution by a
self-consistent mapping onto an effective impurity problem which,
however, must be treated numerically. Neglecting the spatial
correlations, the same method can be applied for an approximate
solution of the Hubbard model in any finite dimension $D<\infty$.
This constitutes the so-called dynamical mean-field theory (DMFT)
\cite{Vol93,GKKR96} which will be employed also for the present
study. It is another intention of the paper to demonstrate that
DMFT can successfully be applied to a film geometry.

The mean-field treatment of the Mott transition in $D<\infty$ rests
on the assumption that spin- and charge fluctuations are reasonably
local. In particular, the electronic self-energy is a local quantity 
within DMFT \cite{MH89b}, $\Sigma_{ij}(E) = \delta_{ij} \Sigma_i(E)$.
The relevance of non-local contributions for a qualitatively correct
description of the Mott transition is not well understood at present. 
Effects of the non-locality of the self-energy can be estimated to 
some extent by conventional second-order $U$ perturbation theory 
(SOPT). As is shown in Ref.\ \cite{SC91}, the local approximation 
is rather well justified for a $D=3$ simple-cubic lattice. Non-local
contributions become more important for $D=2$ and especially for 
$D=1$. With respect to the film geometry, the question is whether
or not non-local contributions can be neglected also near the 
film surfaces. This has been investigated recently by means of
SOPT applied to semi-infinite simple-cubic lattices \cite{PN97c}.
The result is that at the surface the local approximation is as
well justified as in the bulk. 

There is an additional 
suppression of the effects of non-local fluctuations at half-filling:
The low-energy electronic structure, 
being relevant for the transition,
is governed by the (real) linear expansion coefficient
$\beta_{ij}=d\Sigma_{ij}(E)/dE|_{E=0}$. 
Since $\mbox{Re} \Sigma_{ij}(E)$ 
is a symmetric function of $E$ in the particle-hole symmetric case
and for nearest neighbors $i$ and $j$, we have 
$\beta_{\langle ij \rangle}=0$. Within SOPT the second 
nearest-neighbor term $\beta_{ij}$ can be shown \cite{PN97c} 
to be smaller by about two orders of magnitude compared with 
the local one.

This shows that - at least for weak coupling - a local self-energy
functional is a reliable approximation. Beyond the weak-coupling 
regime, however, the approximate locality actually is an assumption 
whose appropriateness has not yet been verified. We nevertheless 
expect DMFT to be a good starting point to study the Mott transition
in a $D=3$ film geometry.

We start our investigations by specifying the film geometries to 
be considered and discuss the DMFT approach with respect to thin
films. In the first part of Sec.~III we briefly focus on the 
Mott transition in the infinitely extended and translationally
invariant model. After that the results for the films are analyzed
in detail. Sec.~IV concludes the study.
\\

{\center \bf \noindent II. DMFT FOR HUBBARD FILMS \\ \mbox{} \\} 

We consider the Hubbard model at half-filling and zero temperature.
Using standard notations, the Hamiltonian reads:
\begin{equation}
  H = \sum_{\langle ij \rangle \sigma} t_{ij} 
  c^\dagger_{i\sigma} c_{j\sigma}
  + \frac{U}{2} \sum_{i\sigma} n_{i\sigma} n_{i-\sigma} \: .
\label{eq:hubbard}
\end{equation}
The hopping integrals $t_{ij}$ are taken to be non-zero between
nearest neighbors $i$ and $j$. $t\equiv-t_{\langle ij \rangle}=1$
sets the energy scale. The thin-film geometry is realized by 
assuming $i$ and $j$ to run over the sites of a system that is 
built up from a finite number $d$ of adjacent layers out of a 
three-dimensional periodic lattice. We study simple-cubic films 
with surface normals along the low-index [100], [110] and [111] 
directions. The model parameters are taken to be uniform, i.~e.\ 
$t$ and $U$ are unchanged at the two film surfaces. With respect 
to the film normal, translational symmetry is broken; lateral 
translational symmetry, however, can be exploited by two-dimensional 
Fourier transformation: The hopping $t_{ij}$ is transformed into
a matrix $\epsilon_{\alpha \beta}({\bf k})$ which is diagonal in 
the wave vector ${\bf k}$ of the first two-dimensional Brillouin
zone (2DBZ). $\alpha, \beta = 1, \dots ,d$ label the different 
layers. For nearest-neighbor hopping and low-index sc films the
Fourier-transformed hopping matrix is tridiagonal with respect to
the layer indices. Its non-zero
elements are given by: 
$\epsilon_\|({\bf k}) \equiv \epsilon_{\alpha \alpha} ({\bf k})$ and
$\epsilon_\perp({\bf k}) \equiv 
\epsilon_{\alpha \alpha\pm 1} ({\bf k})$.
Let $a$ denote the lattice constant. For a sc(100) film the lateral
and normal dispersions then read:
\begin{eqnarray}
  && \epsilon_\|({\bf k}) = 2 t (\cos(k_xa) + \cos(k_ya)) \: ,
  \nonumber \\ &&
  \epsilon_\perp({\bf k})^2 = t^2 \: .
\label{eq:epsilon100}
\end{eqnarray}
Only the square of $\epsilon_\perp({\bf k})$ will enter the physical
quantities we are interested in. For the sc(110) geometry we have:
\begin{eqnarray}
  && \epsilon_\|({\bf k}) = 2 t \cos(k_xa) \: ,
  \nonumber \\ &&
  \epsilon_\perp({\bf k})^2 = 2 t^2 + 2 t^2 \cos(\sqrt{2} k_y a) \: ,
\label{eq:epsilon110}
\end{eqnarray}
and the dispersions for the sc(111) films are:
\begin{eqnarray}
  && \epsilon_\|({\bf k}) = 0 \: ,
  \nonumber \\ &&
  \epsilon_\perp({\bf k})^2 = 3 t^2 + 
  2 t^2 \cos(\sqrt{2} k_y a) 
  \nonumber \\ && \mbox{} \hspace{14mm}
  + 4 t^2 \cos(\sqrt{3/2} k_x a) \cos(\sqrt{1/2} k_y a) \: .
\label{eq:epsilon111}
\end{eqnarray}
Note that the parallel dispersion vanishes since there are no
nearest neighbors within the same layer. Due to the finite film 
thickness $d$, the local free ($U=0$) density of states acquires a 
layer dependence: $\rho^{(0)}_i(E) = \rho^{(0)}_\alpha(E)$ for
sites $i$ within layer $\alpha$. Particle-hole symmetry requires
$\rho_\alpha^{(0)}(E)$ to be a symmetric function of the energy 
for each layer: On the bipartite lattice the odd moments 
$\int E^{2l+1} \rho_\alpha^{(0)}(E) dE$ ($l=0,1,\dots$) 
vanish for all $\alpha$
(cf.\ Ref.\ \cite{PN97c}).

To study the transition from a paramagnetic metal at weak coupling
to a paramagnetic Mott-Hubbard insulator at strong $U$, we restrict
ourselves to the spin-symmetric and (laterally) homogeneous solutions 
of the mean-field equations. As usual \cite{Geb97} we thereby ignore
antiferromagnetic ordering which is expected to be realized in 
the true ground state at any $U>0$. The on-site Green function
$G_{ii}(E) = \langle \langle c_{i\sigma} ; c^\dagger_{i\sigma}
\rangle \rangle_E$ thus depends on the layer index only:
$G_{ii}(E) = G_\alpha(E)$. The same holds for the self-energy
$\Sigma_{ii}(E) = \Sigma_\alpha(E)$ which is a local quantity
within the mean-field approach. Via two-dimensional Fourier
transformation we obtain from the Dyson equation:
\begin{eqnarray}
  G_\alpha(E) &=& \frac{1}{N_\|} \sum_{\bf k} R^{-1}_{\alpha \alpha}
  ({\bf k},E) \nonumber \\
  R_{\alpha \beta}({\bf k},E) &=& (E+\mu) \delta_{\alpha \beta} 
  - \epsilon_{\alpha \beta}({\bf k})
  - \delta_{\alpha \beta} \Sigma_\alpha(E) \: , \nonumber \\
\label{eq:dyson}
\end{eqnarray}
where $N_\|$ is the number of sites within each layer ($N_\| \mapsto
\infty$) and ${\bf k} \in \mbox{2DBZ}$. For the particle-hole
symmetric case, the Fermi energy is given by $\mu=U/2$. Since 
$\epsilon_{\alpha\beta}({\bf k})$ is tridiagonal, the matrix 
inversion is readily performed numerically by evaluating a 
continued fraction of finite depth which is given by the film 
thickness $d$ (cf.\ Ref.\ \cite{Hay80}).

The layer-dependent self-energy $\Sigma_\alpha(E)$ shall be 
calculated within the dynamical mean-field theory (DMFT) 
\cite{Vol93,GKKR96}. This non-perturbative approach treats the
local spin and charge fluctuations exactly. Neglecting spatial
correlations, a {\em homogeneous} lattice problem can be mapped
onto an effective impurity problem supplemented by a self-consistency
condition \cite{GK92a,Jar92}. For the present case of a thin
Hubbard film we have to account for the non-equivalence of sites
within different layers. Therefore, the film problem is mapped
onto a set of $d$ different impurity models with $d$ self-consistency
conditions. 

The DMFT equations are solved by means of the following iterative
procedure: We start with a guess for the self-energy 
$\Sigma_\alpha(E)$. This yields the on-site Green function
$G_\alpha(E)$ via Eq.\ (\ref{eq:dyson}). In the next step we
consider a single-impurity Anderson model (SIAM) \cite{And61}
for each layer $\alpha$:
\begin{eqnarray}
  H^{(\alpha)}_{\rm imp} 
  & \!\!\! = \!\!\! & \sum_\sigma \epsilon_d c^\dagger_\sigma c_\sigma
  + U n_\uparrow n_{\downarrow}
  + \sum_{\sigma, k=2}^{n_s} \epsilon_k^{(\alpha)} 
  a^\dagger_{k\sigma} a_{k\sigma}  
  \nonumber \\ 
  & \!\!\! + \!\!\! & \sum_{\sigma, k=2}^{n_s} \left( V_k^{(\alpha)} 
  a^\dagger_{k\sigma} c_\sigma + \mbox{H.c.} \right) \: .
\label{eq:siam}
\end{eqnarray}
Here $\epsilon_k^{(\alpha)}$ and $V_k^{(\alpha)}$ denote the 
conduction-band energies and the hybridization strengths of the
$\alpha$-th SIAM, respectively. It is sufficient to fix the free
($U=0$) impurity Green function $G^{(0)}_\alpha(E)$ which is 
obtained from the $\alpha$-th DMFT self-consistency condition as:
\begin{equation}
  G_{\alpha}^{(0)}(E) = 
  \left( G_{\alpha}(E)^{-1} + \Sigma_\alpha(E) \right)^{-1} \: .
\label{eq:dmft}
\end{equation}
The crucial step is the solution of the impurity models for 
$\alpha=1,\dots,d$ to get the impurity self-energy $\Sigma_\alpha(E)$ 
which is required for the next cycle.

The computational effort needed for the solution of the impurity
models scales linearly with the system size. It is enhanced by a
factor $d/2$ compared with DMFT applied to the translationally 
invariant (bulk) Hubbard model if one takes into account the mirror 
symmetry with respect to the central layer of the film. Compared 
with the translationally invariant problem, we only found a slight
increase in the number of cycles necessary for the convergence of 
the iterative procedure. The coupling between the different impurity 
problems via their respective baths of conduction electrons turns 
out to be weak.

The application of DMFT to the Hubbard model in thin-film geometry
rests on exactly the same assumption that is necessary for the 
application of DMFT to any finite-dimensional system, namely on
the local approximation for the self-energy. To be precise: the 
self-energy is taken to be local, $\Sigma_{ij}(E) = \delta_{ij} 
\Sigma_i(E)$, and to be given by the (diagrammatic) functional of 
the full {\em local} propagator $G_{ii}(E)$ only. This is sufficient 
to establish the mapping onto the impurity models and to derive the 
self-consistency condition. Since the local approximation is the
only approximation used so far, the hopping between the layers is
treated on the same level as the hopping within each layer. Near
the film center and in the limit of infinite film thickness, our
approach thus recovers the $D=3$ (sc) bulk properties,
{\em irrespective} of the particular film surface geometry.
This represents a non-trivial check of the numerics.

For the solution of the SIAM we employ the exact-diagonalization
(ED) method of Caffarel and Krauth \cite{CK94}. The ED has proven 
its usefulness in a number of previous applications 
\cite{CK94,SRKR94,RMK94,LGK94,GKKR96}. The main idea is to
consider a SIAM with a {\em finite} number of sites $n_s$. This 
results in an obvious shortcoming of the method: ED is not able 
to yield a smooth density of states. Another disadvantage consists
in the fact that finite-size effects are non-negligible whenever
there is a small energy scale relevant for the problem to be 
investigated. With respect to the Mott transition, finite-size 
effects become important close to the critical interaction. On 
the other hand, there are a number of advantages: The ED method 
is based on a simple concept; it is easy to handle numerically 
and computationally fast if compared with the quantum Monte-Carlo 
(QMC) approach \cite{HF86,Jar92,RZK92,GK92b,JP93}. This is of
crucial importance for a systematic study that covers a large
parameter space. Opposed to QMC, the ED method is particularly 
well suited to study the model at $T=0$.

Within the ED algorithm the functional equations for
the self-energy are solved on a discrete mesh on the imaginary
energy axis: $iE_n = i(2n+1) \pi / \widetilde{\beta}$ 
($n=0,1, \dots ,n_{\rm max}$). The {\em fictitious} inverse 
temperature $\widetilde{\beta}$ defines a low-energy cutoff,
$n_{\rm max}$ represents a high-energy cutoff. Eq.\ (\ref{eq:dmft}) 
provides the bath Green function $G_{\alpha}^{(0)}(iE_n)$ from 
which we have to fix the parameters of a SIAM with $n_s$ sites. 
Following Ref.\ \cite{CK94}, this is achieved by minimization of 
the cost function
\begin{equation}
  \chi^2 = 
  \frac{1}{n_{\rm max}+1} 
  \sum_{n=0}^{n_{\rm max}}
  \left|
  G_{\alpha}^{(0)}(iE_n)^{-1} - G_{\alpha, n_s}^{(0)}(iE_n)^{-1}
  \right| 
\label{eq:chi2}
\end{equation}
with respect to the conduction-band energies $\epsilon_k^{(\alpha)}$ 
and the hybridization strengths $V_k^{(\alpha)}$ ($k=2,...,n_s$). 
Thereby, $G_{\alpha}^{(0)}(iE_n)$ is approximated by the free ($U=0$)
Green function of an $n_s$-site SIAM:
\begin{equation}
   G_{\alpha, n_s}^{(0)}(iE_n - \mu)^{-1} = 
   iE_n - \epsilon_d - \sum_{k=2}^{n_s} \frac{(V_{k}^{(\alpha)})^2}
   {iE_n - \epsilon_{k}^{(\alpha)}} \: .
\end{equation}
Obviously, the method is exact for $n_s \mapsto \infty$ only. The 
convergence with respect to $n_s$, however, has been found to be 
exponentially fast \cite{CK94}. Typically $n_s=6-10$ sites are 
sufficient for interaction strengths not too close to the critical
interaction. The results are independent of the low-energy cutoff 
provided that $\widetilde{\beta}^{-1}$ is chosen to be sufficiently 
small. Errors show up in the critical region close to the transition. 
Compared with the error due to the finite $n_s$, however, this is
negligible. Once the SIAM is specified, Lancz\`os technique 
\cite{Hay80} may be employed to calculate the ground state and the 
$T=0$ impurity Green function $G_{\alpha}(iE_n)$. 
The local self-energy of the $\alpha$-th layer can be derived 
from the impurity Dyson equation $\Sigma_\alpha(iE_n) = 
G_{\alpha}^{(0)}(iE_n)^{-1} - G_{\rm imp}(iE_n)^{-1}$. 
\\

{\center \bf \noindent III. RESULTS AND DISCUSSION \\ \mbox{} \\} 

{\center \bf \noindent A. Bulk \\ \mbox{} \\} 

Let us first concentrate on the $T=0$ Mott transition in the 
translationally invariant Hubbard model before we come to the 
discussion of the results for the film geometry. This case has 
been the subject of numerous DMFT studies during the recent years 
\cite{GKKR96,GK92a,CK94,RMK94,LGK94,GK93,RKZ94,MSK+95,BHP98}. 
Most investigations refer to the Bethe lattice with infinite 
connectivity where we have a semi-elliptical free density of 
states. However, within DMFT no qualitative changes are expected 
when considering the $D=3$ simple-cubic lattice which also yields 
a symmetric and bounded free density of states. The $D=3$ sc lattice 
is considered here since it represents the limit of infinite film 
thickness $d\mapsto \infty$ with respect to our film results. 

\begin{figure}[t] 
\vspace{-4mm}
{\psfig{figure=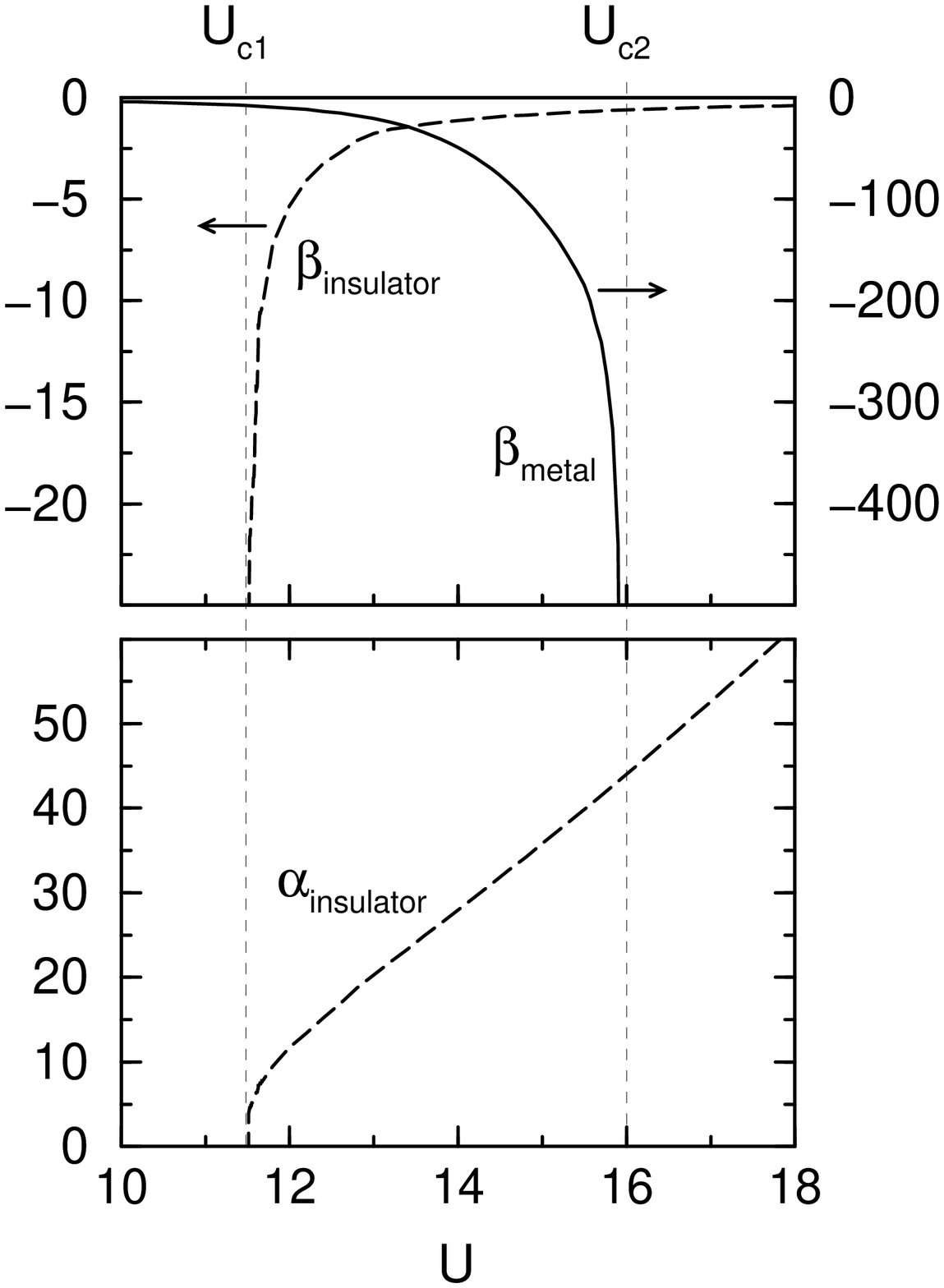,width=85mm,angle=0}}
\vspace{-16mm}

\parbox[]{85mm}{\small Fig.~1.
Results for the (bulk) Mott transition in the half-filled Hubbard
model as obtained within DMFT-ED. $U$ dependence of the linear
coefficient $\beta$ in the low-energy expansion of $\Sigma(E)$
for the metallic (solid line) and the insulating solution (dashed 
line). The critical interactions $U_{c1}$ and $U_{c2}$ are indicated 
by the vertical lines. Lower panel: $1/E$ coefficient $\alpha$ in 
the low-energy expansion for the insulating solution. Calculation 
for a $D=3$ simple-cubic lattice. Nearest-neighbor hopping $t=1$. 
Width of free density of states: $W=12$. $n_s=8$.
}
\label{fig:bulksc}
\end{figure}

Fig.~1 shows the $U$ dependence of the coefficients $\alpha$ and 
$\beta$ in the low-energy Laurent expansion of the self-energy: 
\begin{equation}
  \Sigma(E) = \frac{\alpha}{E} + \frac{U}{2} + \beta E + \cdots \: .
\end{equation}
The calculation is performed for $n_s=8$. For a metal, i.~e.\ if 
$\alpha=0$, the coefficient $\beta$ yields the usual quasi-particle 
weight $z=(1-\beta)^{-1}$ \cite{MH89c}. We find a metallic 
solution for interaction strengths up to a critical value
$U_{c2}=16.0$ ($U_{c2}=4W/3$ in terms of the width of the free 
density of states $W=12$). As $U$ approaches $U_{c2}$, $\beta$ 
diverges, i.~e.\ the quasi-particle weight vanishes. At $U_{c2}$
the metallic solution continuously coalesces with the insulating
solution that is found for strong $U$. For decreasing $U$ the 
insulating phase ceases to exist below another critical interaction 
strength $U_{c1} = 11.5$ which is marked by the vanishing $1/E$ 
expansion coefficient $\alpha$ as well as by $\beta \mapsto -\infty$ 
(Fig.~1). We find $U_{c1} < U_{c2}$; there is a region where both, 
the metallic and the insulating solution, coexist \cite{CKcomment}. 

A precise determination of the critical interactions, at least 
of $U_{c2}$, is not possible by means of the ED method (see next 
section). Qualitatively, however, the results are consistent with 
the findings for the $D=\infty$ Bethe lattice 
\cite{GK92a,CK94,RMK94,LGK94,GK93,RKZ94,MSK+95,BHP98}:
Within the iterative perturbation theory (IPT) \cite{GK92a,GKKR96} 
a narrow quasi-particle resonance is seen to develop at the Fermi
energy for increasing $U$ in the metallic solution. The spectrum 
has a three-peak structure, two additional charge-excitation peaks 
(Hubbard bands) show up at $E\approx \pm U/2$. For $U\mapsto U_{c2}$
the effective mass $z^{-1}$ diverges as in the Brinkman-Rice 
variational approach \cite{BR70}. On the other hand, at strong 
$U$ the Hubbard bands are well separated by a gap in the insulating 
solution as in the Hubbard-III approach \cite{Hub64b}. One can
follow up the insulating solution by decreasing $U$ down to 
$U=U_{c1}$. For $T=0$ a coexistence of the solutions ($U_{c1} 
< U_{c2}$) has first been observed within IPT \cite{GK93,RKZ94}. 
It is confirmed by ED \cite{RMK94} as well as by means of a recent 
numerical renormalization-group calculation (NRG) \cite{BHP98,bulla2}
which is particularly suited to study the critical regime. The 
comparison between the respective internal energies within IPT 
\cite{GKKR96}, ED \cite{RMK94,GKKR96} and NRG \cite{bulla2} as 
well as a simple argument mentioned in Ref.\ \cite{MSK+95}
show that the metallic phase is stable against the insulating 
one in the whole coexistence region up to $U_{c2}$: The $T=0$ 
transition is found to be of second order. 

One should also be aware of serious physical arguments 
\cite{LN98,Geb97} which have been raised against a transition 
scenario with two different critical interaction strengths.
The exhaustion problem mentioned 
in Ref.\ \cite{LN98} at least shows that a conclusive understanding 
of the Mott transition in $D=\infty$ has not yet been achieved. From 
the above discussion and our own results, however, we can conclude 
that the numerical evidences for the existence of 
a finite coexistence region are 
strong. Another problem is tackled in Ref.\ \cite{Keh98}, where the 
concept of a preformed gap \cite{GKKR96} is shown to be at variance 
with Fermi liquid theory. Our ED study cannot contribute to settle 
this interesting question 
since the detailed picture of
the low-energy electronic structure in the limit $U\mapsto U_{c2}$
is concerned.
\\

{\center \bf \noindent B. Films \\ \mbox{} \\} 

\begin{figure}[b] 
\vspace{2mm}
\centerline{\psfig{figure=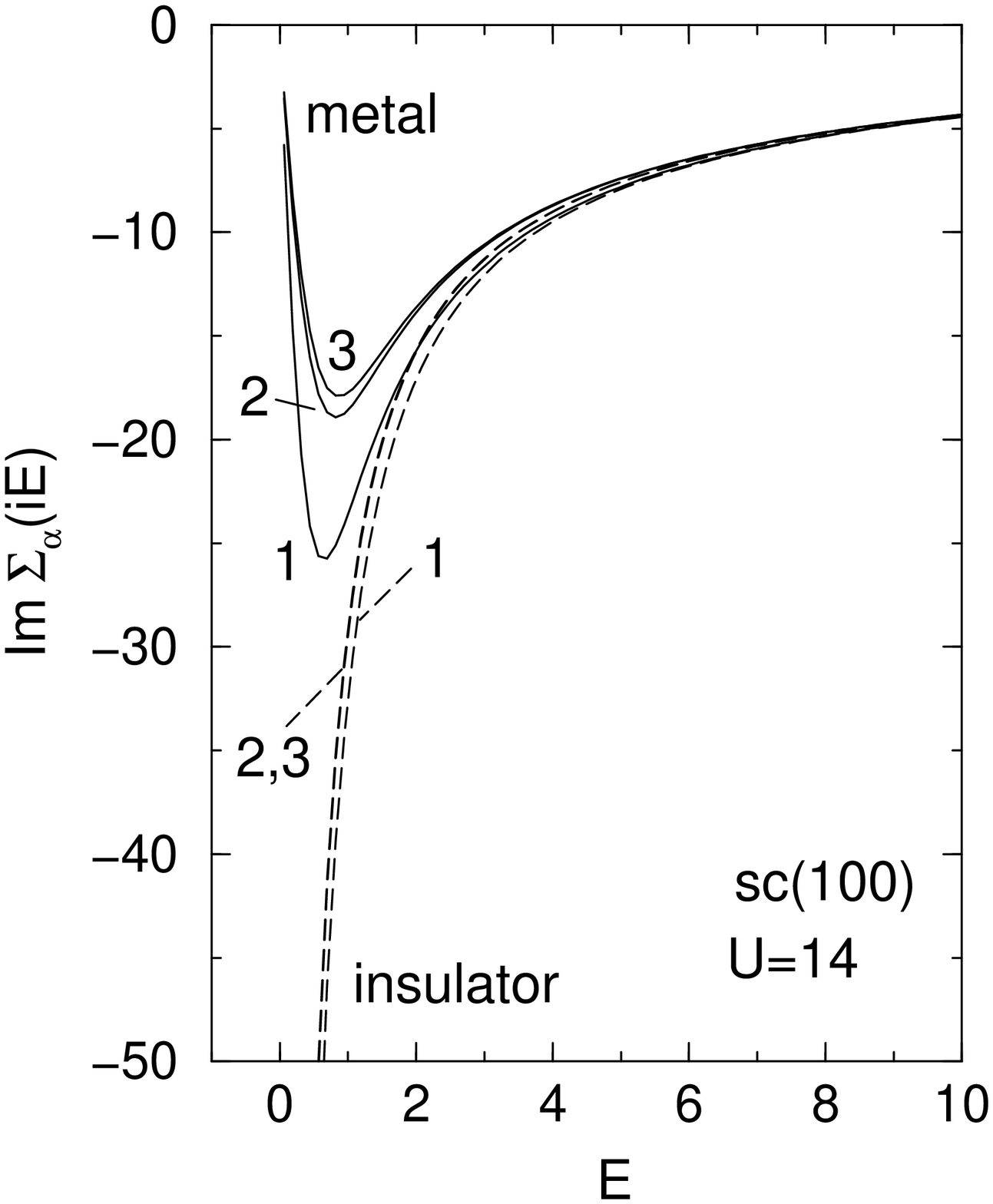,width=75mm,angle=0}}
\vspace{3mm}

\parbox[]{85mm}{\small Fig.~2.
Imaginary part of the layer-dependent self-energy $\Sigma_\alpha$
on the discrete mesh of the imaginary energies 
$iE_n=(2n+1)\pi/\widetilde{\beta}$ ($\widetilde{\beta}=50$, 
$\widetilde{\beta}^{-1} = 0.0016 \, W$) for
the $d=5$ sc(100) film at $U=14$. 
Results for the three inequivalent layers $\alpha=1-3$ as indicated.
$\alpha=1$: surface layer.
Solid lines: metallic phase.
Dashed lines: metastable insulating phase.
\label{fig:smetins}
}
\end{figure}

\begin{figure}[t] 
\vspace{2mm}
\centerline{\psfig{figure=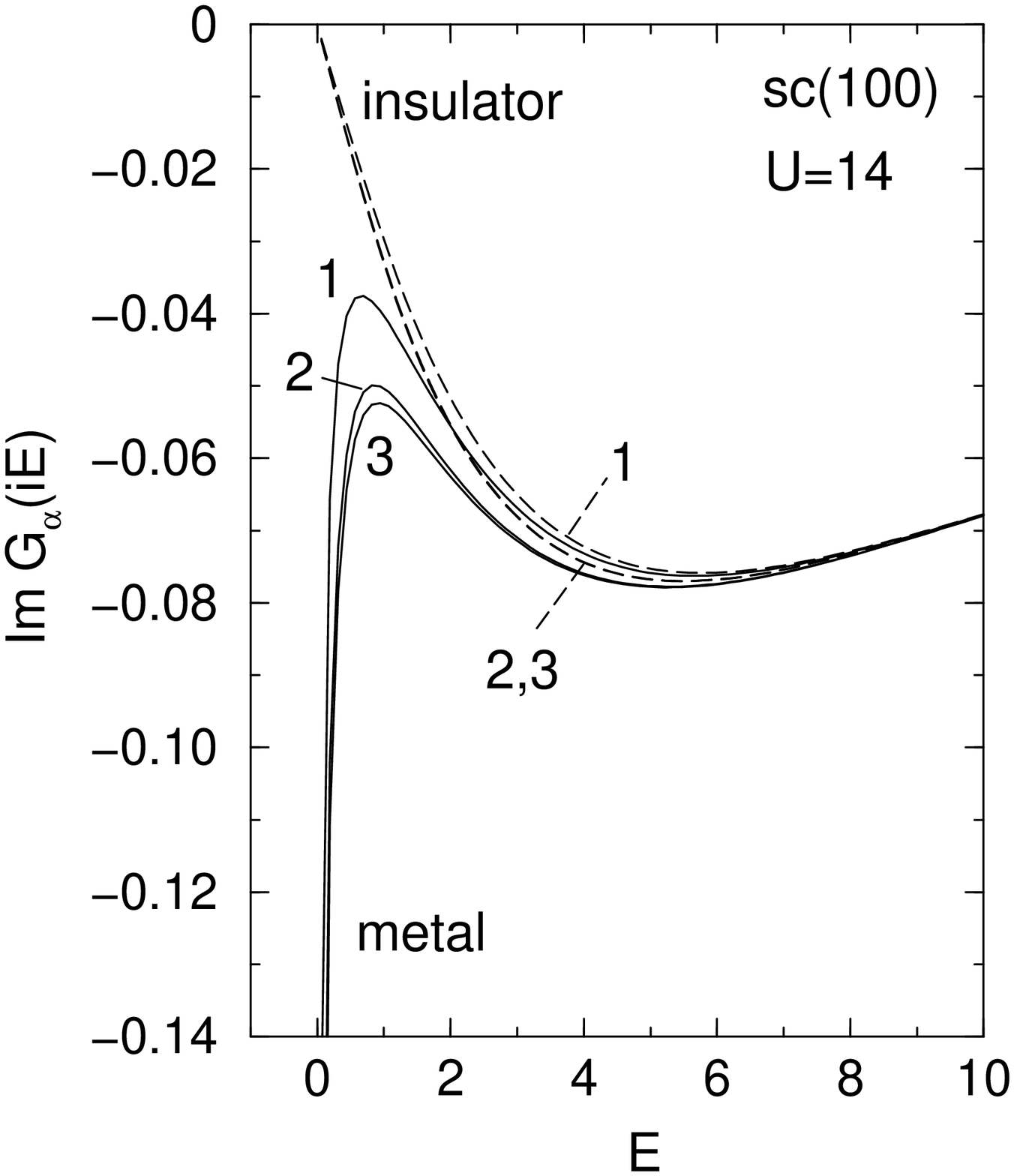,width=75mm,angle=0}}
\vspace{3mm}

\parbox[]{85mm}{\small Fig.~3.
The imaginary part of the Green functions corresponding to the 
self-energies in Fig.~2.
\label{fig:gmetins}
}
\end{figure}

\begin{figure}[b] 
\vspace{-2mm}
\center{\psfig{figure=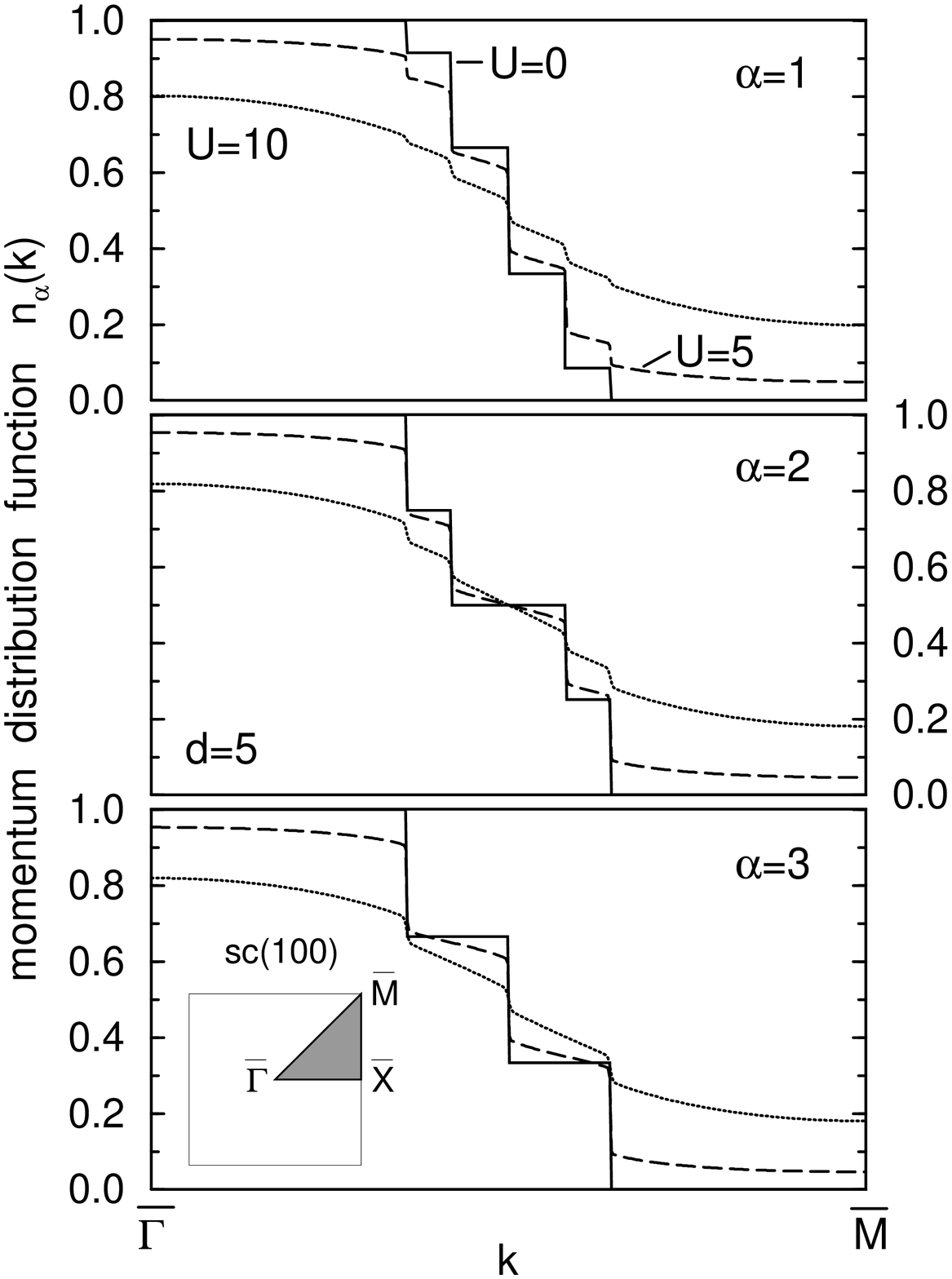,width=85mm,angle=0}}
\vspace{-6mm}

\parbox[]{85mm}{\small Fig.~4.
Momentum distribution function $n_\alpha({\bf k})$ for the
(non-equivalent) layers $\alpha=1,2,3$ 
along a high-symmetry direction in the irreducible part of the 2DBZ.
Results for the $d=5$ sc(100) film
at different $U$.
\label{fig:nofk}
}
\end{figure}

In the following we discuss our results obtained by the ED approach
to investigate the characteristics of the Mott transition in thin 
Hubbard films.
Routinely, the calculations for the Hubbard films have been 
performed with $n_s=8$ sites in the effective impurity problems. 
We systematically compared with $n_s=6$ and also checked against 
$n_s=10$ at a few data points. It turns out that there are no 
significant differences between the results for the different 
$n_s$ as long as $U$ is not too close to $U_{c2}$ (see Fig.~10 and 
the related discussion). Choosing a small fictitious temperature 
$\widetilde{\beta}^{-1} = 0.0016 \, W$ and a large high-energy 
cutoff $(2n_{\rm max} - 1)\pi/\widetilde{\beta} > 2U$ ensures 
the results to be independent of the discrete energy mesh (see
also discussion of Fig.~10). For the electron-hole symmetric case 
we can reduce the number of parameters in the multi-dimensional 
minimization [Eq.\ (\ref{eq:chi2})] and for the DMFT self-consistency 
[Eq.\ (\ref{eq:dmft})] by setting $\epsilon_{k}=-\epsilon_{k'}$ and 
$V^2_{k}=V^2_{k'}$ for $k,k'$ with $k+k'=n_s+2$. For $n_s=8$ and 
$d=6$, which are typical values considered here, there are 
$(n_s-1)d/2=21$ parameters to be determined. The stabilization 
of paramagnetic solutions has turned out to be unproblematic.
A mixing of ``old'' and ``new'' parameters (50\%), however, has 
been found to be necessary to obtain a converging self-consistency
cycle for the sc(111) films. Apart from the coexistence of a metallic
and an insulating solution in a certain $U$ range, we always found
a unique solution of the mean-field equations.

Fig.~2 shows the imaginary part of the self-energy as obtained for
a $d=5$ sc(100) film. At $U=14$ we find two solutions. In the 
metallic one there is a significant layer dependence of 
$\mbox{Im} \, \Sigma_\alpha(iE)$ with a considerably larger slope
$d\Sigma_\alpha/dE$ for $E\mapsto 0$ in the surface layer $(\alpha=1)$.
This is plausible since the surface-layer sites have a reduced
coordination number $n$ resulting in a diminished variance
$\Delta = n t^2$ of the free local density of states. Thus
$U/\sqrt{\Delta}$ is larger compared with the film center tending
to enhance correlation effects. The large values of $d\Sigma_\alpha/dE$
indicate that the system is close to the transition.

At $E=0$ the imaginary part of the self-energy vanishes for all
layers as it must be for a Fermi liquid (note that in Fig.~2
$\mbox{Im} \, \Sigma$ is plotted on the discrete mesh $iE_n$ only).
A weak layer dependence is noticed for the insulating solution.
In this case $\mbox{Im} \, \Sigma_\alpha(iE)$ diverges for 
$E\mapsto 0$.

The imaginary parts of the corresponding Green functions are shown
in Fig.~3. Again, the layer dependence is more pronounced for the
metallic solution. For the insulating solution we have 
$\mbox{Im} \, G_\alpha(i0^+)=0$, i.~e.\ a vanishing layer density of 
states at $E=0$. Contrary, $\mbox{Im} \, G_\alpha(i0^+)$ stays finite
for the metal. The $E=0$ value is given by Eq.\ (\ref{eq:dyson})
where
\begin{eqnarray}
  R_{\alpha \beta} ({\bf k},i0^+) & = &
  (\mu +i0^+) 
  \delta_{\alpha \beta} - \epsilon_{\alpha \beta}({\bf k})
  - \delta_{\alpha \beta} \Sigma_\alpha(0)
  \nonumber \\
  & = & i0^+ \delta_{\alpha \beta} - \epsilon_{\alpha \beta}({\bf k})
  = R_{\alpha \beta} ({\bf k},i0^+)\Big|_{U=0} \: ,
  \nonumber \\
\end{eqnarray}
and where we used that $\mu = \Sigma_\alpha(0) = U/2$ for all 
layers $\alpha$. This implies that (for a local Fermi liquid) 
the layer density of states at the Fermi energy $\rho_\alpha(0) 
= - \mbox{Im} \, G_\alpha(i0^+)/\pi$ is unrenormalized by the 
interaction. As for $U=0$, however, it is layer dependent. For 
the translationally invariant Hubbard model with local self-energy 
the pinning of the density of states is a well known and general 
consequence of Luttinger's sum rule \cite{MH89c}. In the case 
of Hubbard films (and within DMFT) there is a pinning of 
$\rho_\alpha(0)$ only for the case of half-filling since off 
half-filling $\Sigma_\alpha(0)$ acquires a layer dependence.

From the low-energy expansion of the self-energy for the metallic
solution we can calculate
the so-called layer-dependent quasi-particle weight
\begin{equation}
  z_\alpha = \left( 1 - \frac{d \Sigma_\alpha(E)}
  {dE} \Big|_{E=0} \right)^{-1} \: .
\label{eq:zq}
\end{equation}
Once a self-consistent solution of the mean-field equations on the
discrete energy mesh $iE_n$ has been obtained, the self-energy can
be determined in the entire complex energy plane, and there is no
difficulty to calculate the derivative in Eq.\ (\ref{eq:zq}).
Let us briefly discuss the physical meaning of $z_\alpha$ which 
for a film geometry is slightly different compared with the bulk 
case \cite{Lut60}. Exploiting the lateral translational symmetry 
and performing a two-dimensional Fourier transformation, the
Green function at low energies is given by:
\begin{eqnarray}
  G_{\alpha \beta}({\bf k},E) & \equiv & \frac{1}{N_\|}
  \sum_{i_\| j_\|} e^{i{\bf k}({\bf R}_{i_\|}-{\bf R}_{j_\|})}
  \langle \langle c_{i_\| \alpha \sigma} ;
  c^\dagger_{j_\| \beta \sigma}  \rangle \rangle
\nonumber \\ 
  &=& \sqrt{z_\alpha} 
  \left( \frac{\bf 1}{E {\bf 1} - {\bf T}({\bf k})} 
  \right)_{\alpha \beta} \sqrt{z_\beta} \: .
\label{eq:gk}
\end{eqnarray}
Here $({\bf T}({\bf k}))_{\alpha \beta} = \sqrt{z_\alpha}
\epsilon_{\alpha \beta}({\bf k}) \sqrt{z_\beta}$ is the renormalized
hopping matrix. $i_\|$ and $j_\|$ label the $N_\|$
sites in the layers $\alpha$ and $\beta$, respectively, and 
${\bf k} \in \mbox{2DBZ}$. Only the linear term in the expansion
of the self-energy is taken into account. For each wave vector
${\bf k}$ the matrix ${\bf T}({\bf k})$ can be diagonalized by
a unitary transformation which is mediated by a matrix 
$U_{\alpha r}({\bf k})$. The $d$ eigenvalues $\eta_r({\bf k})$
of ${\bf T}({\bf k})$ ($r=1,\dots, d$) yield the dispersions of 
the quasi-particle bands near the Fermi energy and determine the
$d$ one-dimensional Fermi ``surfaces'' in the 2DBZ via
$\eta_r({\bf k})=0$. As can be seen from Eq.\ (\ref{eq:gk}),
when ${\bf k}$ approaches the $r$-th Fermi surface there is a
discontinuous drop of the $\alpha$-th momentum-distribution
function
\begin{equation}
  n_\alpha({\bf k}) = -\frac{1}{\pi} \int_{-\infty}^0 
  \mbox{Im} \, G_{\alpha \alpha}({\bf k},E+i0^+) \, dE
  \: ,
\end{equation}
which is given by:
\begin{equation}
  \delta n_\alpha({\bf k}_{\rm F}^r) = 
  \big| U_{\alpha r}({\bf k}_{\rm F}^r) \big|^2
  \cdot z_\alpha
  \: .
\end{equation}
Summing over the $d$ Fermi surfaces, we have:
\begin{equation}
  z_\alpha = \sum_{r=1}^d \delta n_\alpha({\bf k}_{\rm F}^r) 
  \: .
\end{equation}

\begin{figure}[t] 
\vspace{-2mm}
\center{\psfig{figure=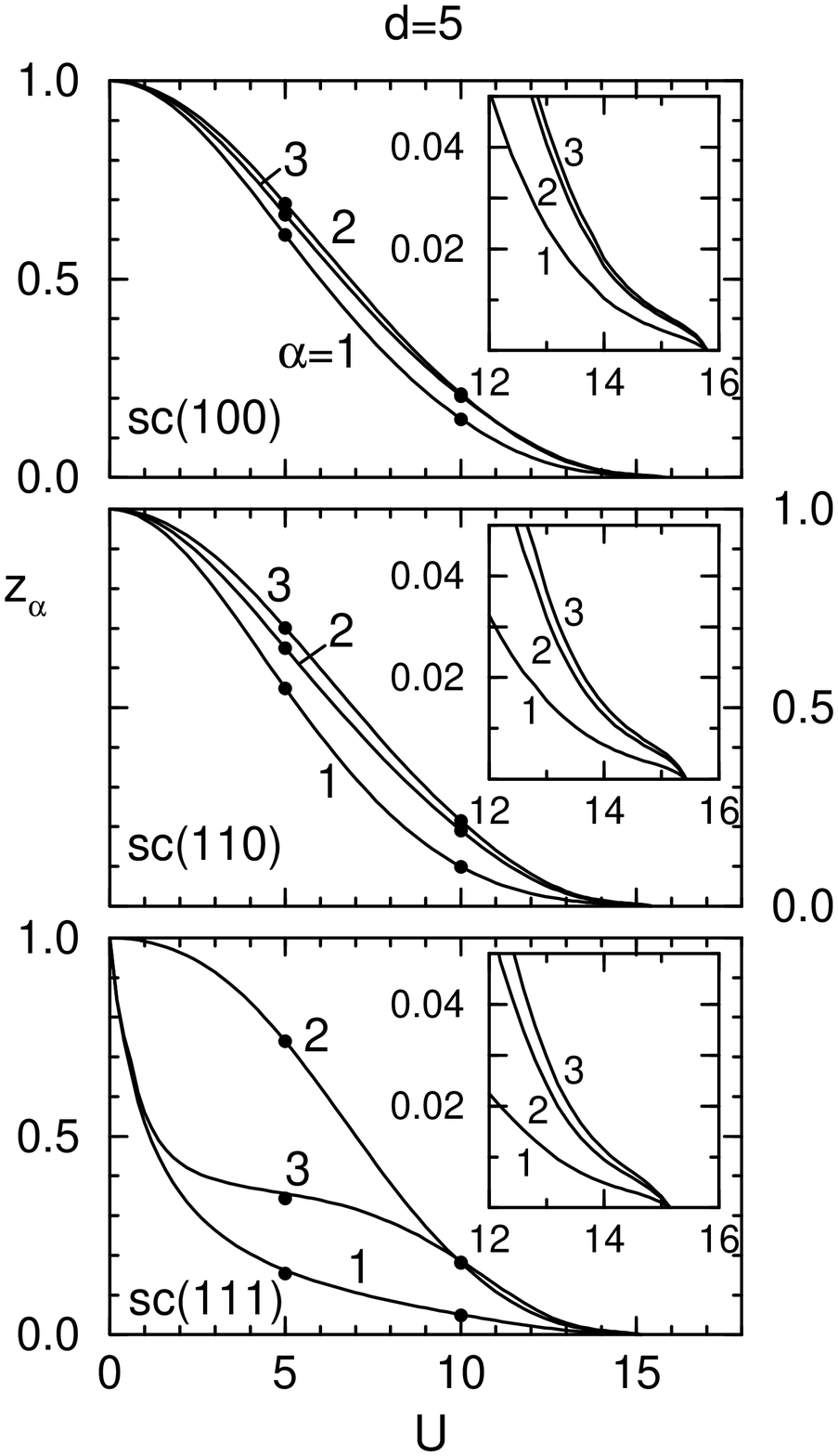,width=90mm,angle=0}}
\vspace{-6mm}

\parbox[]{85mm}{\small Fig.~5.
Layer-dependent quasi-particle weight $z_\alpha$ as a function of
$U$ for a $d=5$ layer film in sc(100), sc(110) and sc(111) geometry.
Calculation for $n_s=8$ (solid lines) and $n_s=10$ (dots). The
insets show $z_\alpha(U)$ in the critical regime.
\label{fig:zd5}
}
\end{figure}

\begin{figure}[t] 
\vspace{-2mm}
\centerline{\psfig{figure=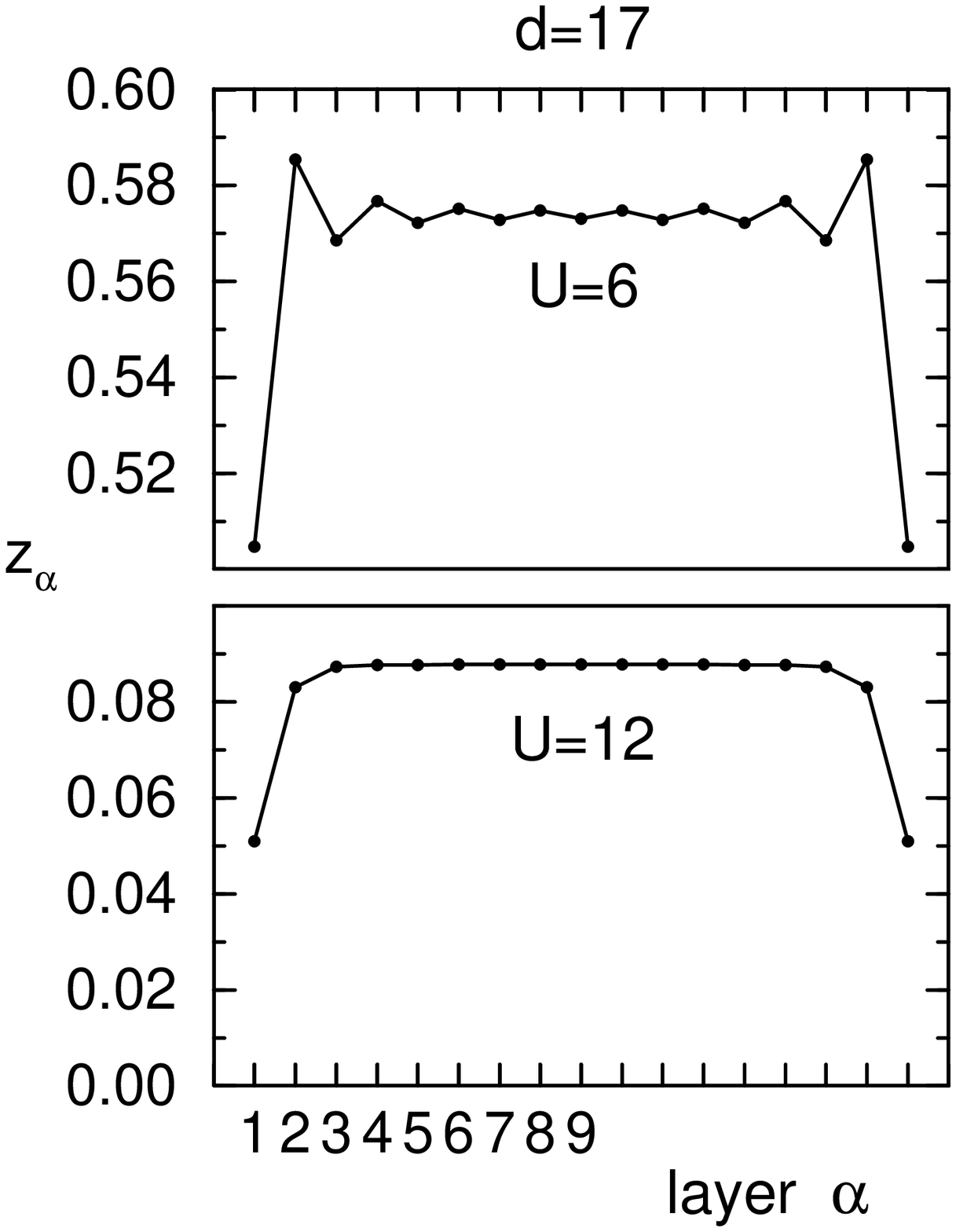,width=80mm,angle=0}}
\vspace{-3mm}

\parbox[]{85mm}{\small Fig.~6.
Layer dependence of the quasi-particle weight for $U=6$ and $U=12$
in the $d=17$ sc(100) film.
\label{fig:zlay}
}
\end{figure}

The momentum-distribution function can be calculated on the 
imaginary energy axis via:
\begin{equation}
  n_\alpha({\bf k}) = \frac{1}{2} + \lim_{\beta \mapsto \infty}
  \frac{2}{\beta} \, \mbox{Re} \sum_{n=0}^\infty 
  G_{\alpha \alpha} ({\bf k}, iE_n) \: .
\end{equation}
Fig.~4 shows $n_\alpha({\bf k})$ along a high-symmetry direction in
the 2DBZ for the $d=5$ sc(100) film. For the non-interacting system
the 5 bands $\eta_r({\bf k})$ are occupied at the 
$\overline{\Gamma} = (0,0)$ point and empty at 
$\overline{M} = (\pi,\pi)$.
Consequently, all bands cross the Fermi energy along the 
$\overline{\Gamma} \overline{M}$ direction, and discontinuinities
can be seen at 5 Fermi wave vectors ${\bf k}_{\rm F}^r$. Due to
symmetries $U_{\alpha r}({\bf k}_{\rm F}^r)$ may vanish in some 
cases. For example, at the Fermi wave vector 
${\bf k}_{\rm F} = (\pi/2,\pi/2)$ we have 
$\epsilon_\|({\bf k}_{\rm F}) = 0$, and one can easily prove
$(1,0,-1,0,1)$ to be an eigenvector of the hopping matrix
$\epsilon_{\alpha \beta}({\bf k}_{\rm F})$ with eigenvalue $\eta=0$.
This implies that there is no discontinuity of $n_{\alpha=2}$
(and $n_{\alpha=4}$) at ${\bf k} = (\pi/2,\pi/2)$.  

For the interacting system, $\delta n_\alpha({\bf k}_{\rm F}^r)$ is
reduced by the layer-dependent factor $z_\alpha < 1$. The Fermi
surfaces themselves, however, remain unchanged since
$\det (\epsilon_{\alpha \beta}({\bf k}) ) = 0$ implies
$\det (\sqrt{z_\alpha} \epsilon_{\alpha \beta}({\bf k}) 
\sqrt{z_\beta}) = 0$ for any ${\bf k}$. Generally speaking, 
the invariance of the Fermi surfaces is a consequence of the
manifest particle-hole symmetry at half-filling and of the
local approximation for the self-energy.

From Eq.\ (\ref{eq:gk}) we have:
\begin{equation}
  z_\alpha = \frac{1}{N_\|} \sum_{\bf k}  
  \int_{-\infty}^\infty -\frac{1}{\pi}  \mbox{Im} \,
  G^{\rm (coh)}_{\alpha \alpha}({\bf k}, E+i0^+) \, dE \: ,
\end{equation}
which shows that $z_\alpha$ also yields the weight of the
coherent quasi-particle peak in the layer-resolved 
density of states. $z_\alpha$ can serve as an 
``order parameter'' for the Mott transition.
For $z_\alpha=0$ the system is a Mott-Hubbard insulator.

Fig.~5 shows $z_\alpha$ as a function of $U$ in the metallic solution
for the sc(100), sc(110) and the sc(111) films with thickness $d=5$. 
The layer-dependent quasi-particle weight decreases from its 
non-interacting value 
$z_\alpha=1$ to $z_\alpha=0$. In the weak-coupling regime there 
is a quadratic dependence $1-z_\alpha(U) \propto U^2$ which is 
consistent with perturbation theory in $U$ \cite{PN97c}. An
almost linear dependence is seen for the sc(111) film for 
$\alpha=1$ and $\alpha=3$ which indicates an early breakdown
of perturbation theory in this case. For each 
film geometry considered, there is a unique critical interaction 
$U_{c2}$ where all functions $z_\alpha(U)$ simultaneously approach 
zero; the whole system can be either metallic or insulating. 

Depending on the film geometry, we notice a weak (sc(100)) or a
rather strong (sc(111)) layer dependence of the quasi-particle
weight. For weak and intermediate couplings we may observe an 
oscillatory dependence on $\alpha$ (sc(100)). For $U$ closer to 
$U_{c2}$, the behavior changes qualitatively; here $z_\alpha$ 
monotonously increases with increasing distance from a film surface.
This is a typical effect which is also observed for still thicker
films. Fig.~6 illustrates this fact. It shows a film profile of 
the quasi-particle weight for a $d=17$ sc(100) film at $U=6$
($z_\alpha$ oscillating) and $U=12$ ($z_\alpha$ monotonous).

\begin{figure}[t] 
\vspace{-18mm}
\center{\psfig{figure=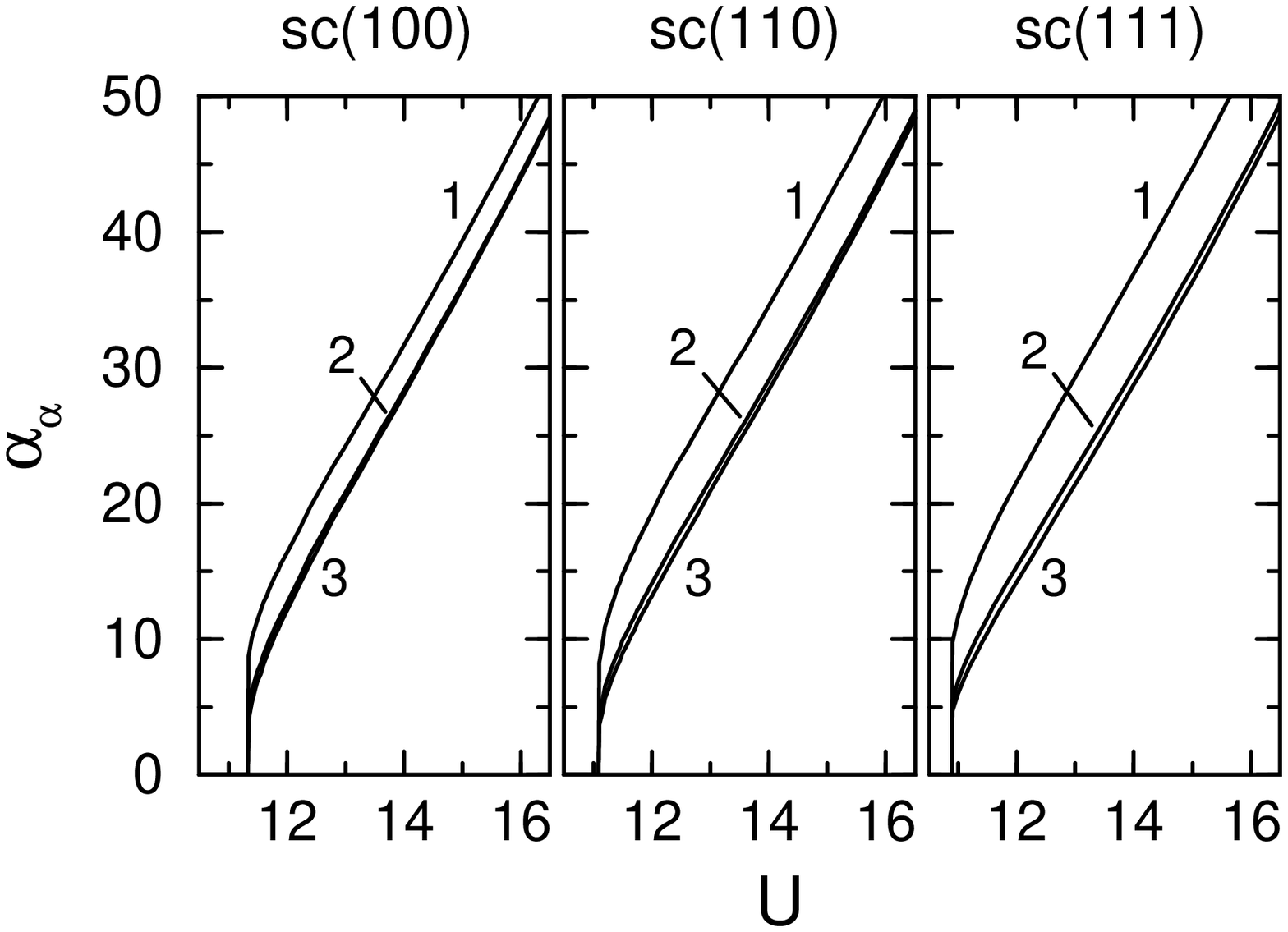,width=80mm,angle=0}}
\vspace{-33mm}

\parbox[]{85mm}{\small Fig.~7.
Layer-dependent $1/E$ (low-energy) coefficient as a function of $U$ 
for the insulating phase and different film geometries. Film 
thickness: $d=5$.
\label{fig:alphau}
}
\end{figure}

\begin{figure}[b] 
\vspace{-7mm}
\center{\psfig{figure=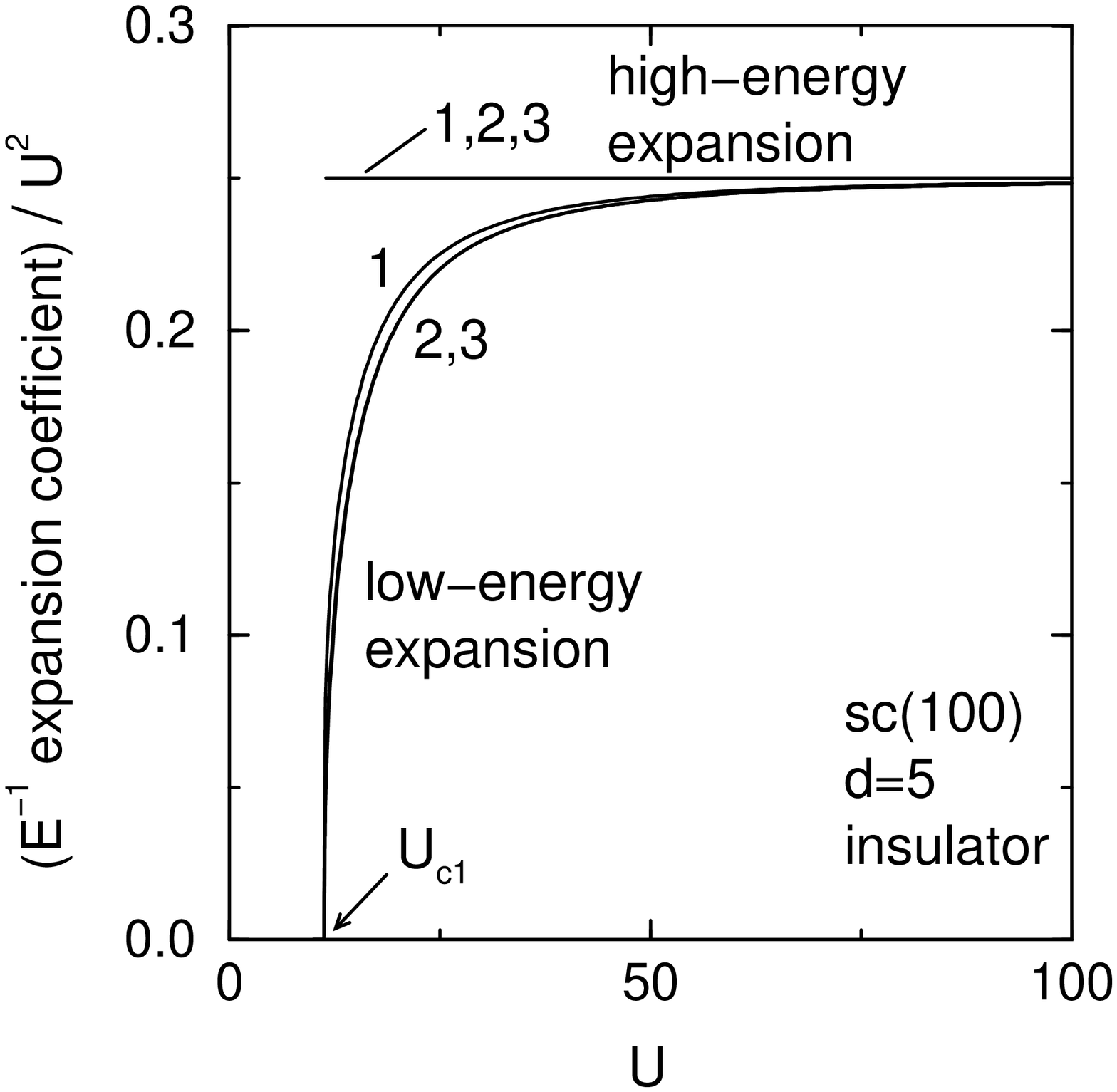,width=80mm,angle=0}}
\vspace{-7mm}

\parbox[]{85mm}{\small Fig.~8.
Layer-dependent $1/E$ coefficient in the low- and the high-energy
expansion of the self-energy as a function of $U$. Insulator on 
the $d=5$ sc(100) film.
\label{fig:isoinfty}
}
\end{figure}

In all cases the quasi-particle weight 
of the surface layer $z_{\alpha=1}$ is significantly 
reduced compared with $\alpha=2,3$.
As has been mentioned above, this 
can be attributed to the lowered film-surface coordination 
number. For $d=5$ (Fig.~5) 
the critical interaction $U_{c2}$ is significantly
reduced compared with the bulk ($d\mapsto \infty$) value.
For the sc(100) film we find
$U_{c2}=15.8$, for the sc(110) $U_{c2}=15.4$, and for the 
sc(111) geometry we have $U_{c2}=15.2$. This has to be compared
with the bulk value $U_{c2}=16.0$ (see Fig.~1). The critical 
interaction is the lower the more open is the film surface. We have 
$n_{(100)}=5$, $n_{(110)}=4$, $n_{(111)}=3$ for the coordination
numbers at the film surfaces while in the film center $n=6$.
(Errors for the $U_{c2}$ values are discussed below, the 
observed trends are not affected).

The same trend is also found for the critical interaction $U_{c1}$
where the insulating solution disappears. Fig.~7 shows the $U$
dependence of the $1/E$ coefficient in the low-energy expansion
of the self-energy. For the sc(100) film we find $U_{c1} = 11.3$
while $U_{c1} = 11.1$ for the sc(110) geometry. With $U_{c1} = 10.9$
the lowest value is observed for the sc(111) film. Contrary to the
quasi-particle weight in the metallic solution, there is always a
monotonous increase of the $1/E$ coefficient when passing from the
film center to one of the surfaces for the insulating solution.
In particular, the surface-layer value is significantly enhanced. 
As $U\mapsto U_{c1}$, the $1/E$ coefficient non-continuously drops
to zero. For the thin-film geometries this is much more apparent
than for the bulk (see Fig.~1).

The almost linear $U$ dependence of the $1/E$ coefficient well 
above $U_{c1}$ (Fig.~7) alters for still higher $U$: The regime 
of very strong Coulomb interaction is shown in Fig.~8
for the sc(100) film. Eventually, for $U\mapsto \infty$, there is
a quadratic dependence of the $1/E$ coefficient on $U$. It smoothly
approaches the quadratic $U$ dependence of the high-energy expansion
of the self-energy: $\Sigma_\alpha(E) = U/2 + U^2 / 4 E + \cdots$ 
\cite{PHN98}. The system behaves more and more atomic-like.

\begin{figure}[t] 
\vspace{-0mm}
\center{\psfig{figure=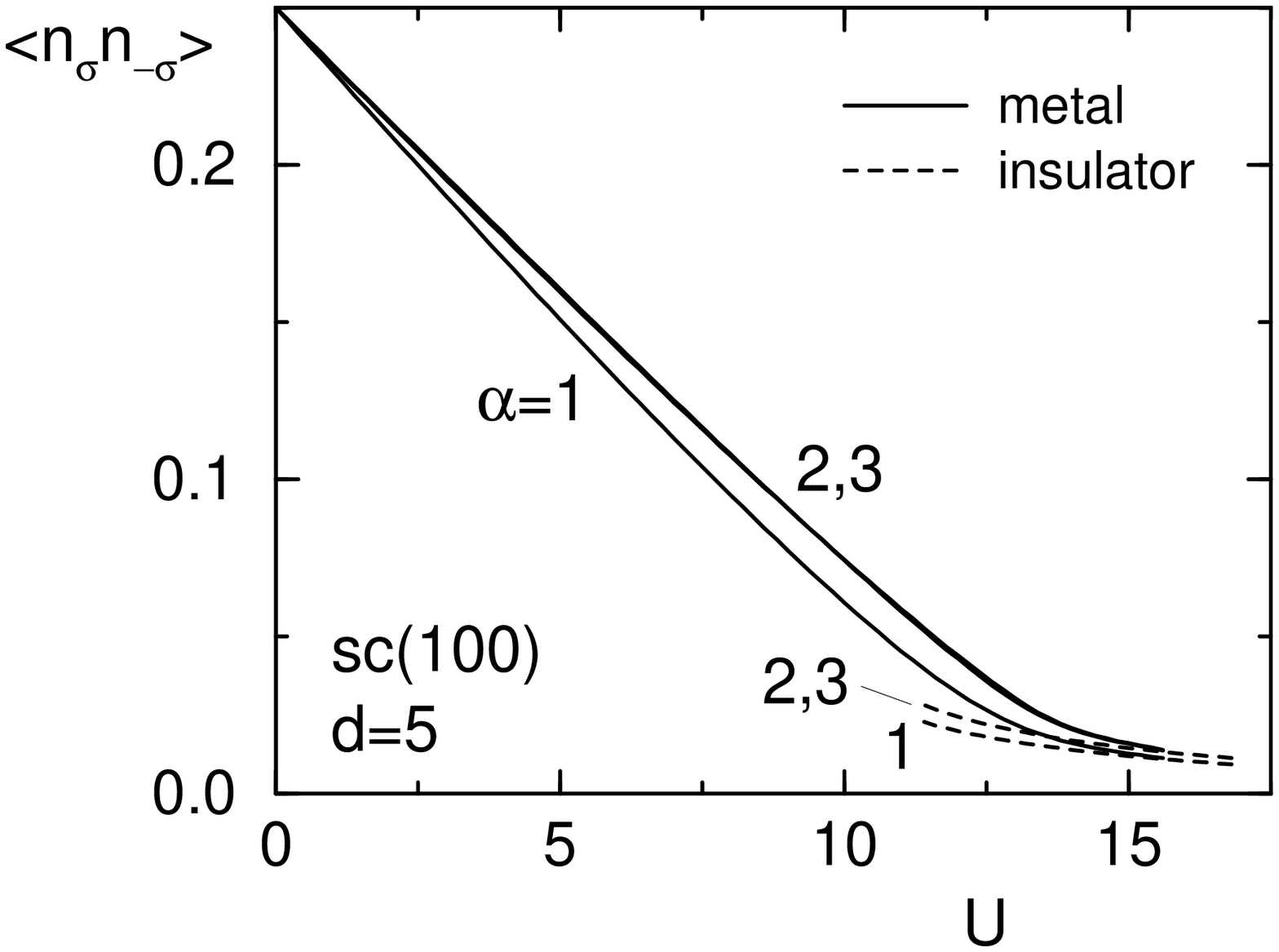,width=80mm,angle=0}}
\vspace{-3mm}

\parbox[]{85mm}{\small Fig.~9.
Layer-dependent double occupancy 
$\langle n_{i\sigma} n_{i-\sigma} \rangle$ (with $i\in \alpha$)
as a function of $U$. Metallic (solid lines) and the insulating 
solution (dashed lines) for the $d=5$ sc(100) film.
\label{fig:double}
}
\end{figure}

\begin{figure}[t] 
\vspace{-2mm}
\center{\psfig{figure=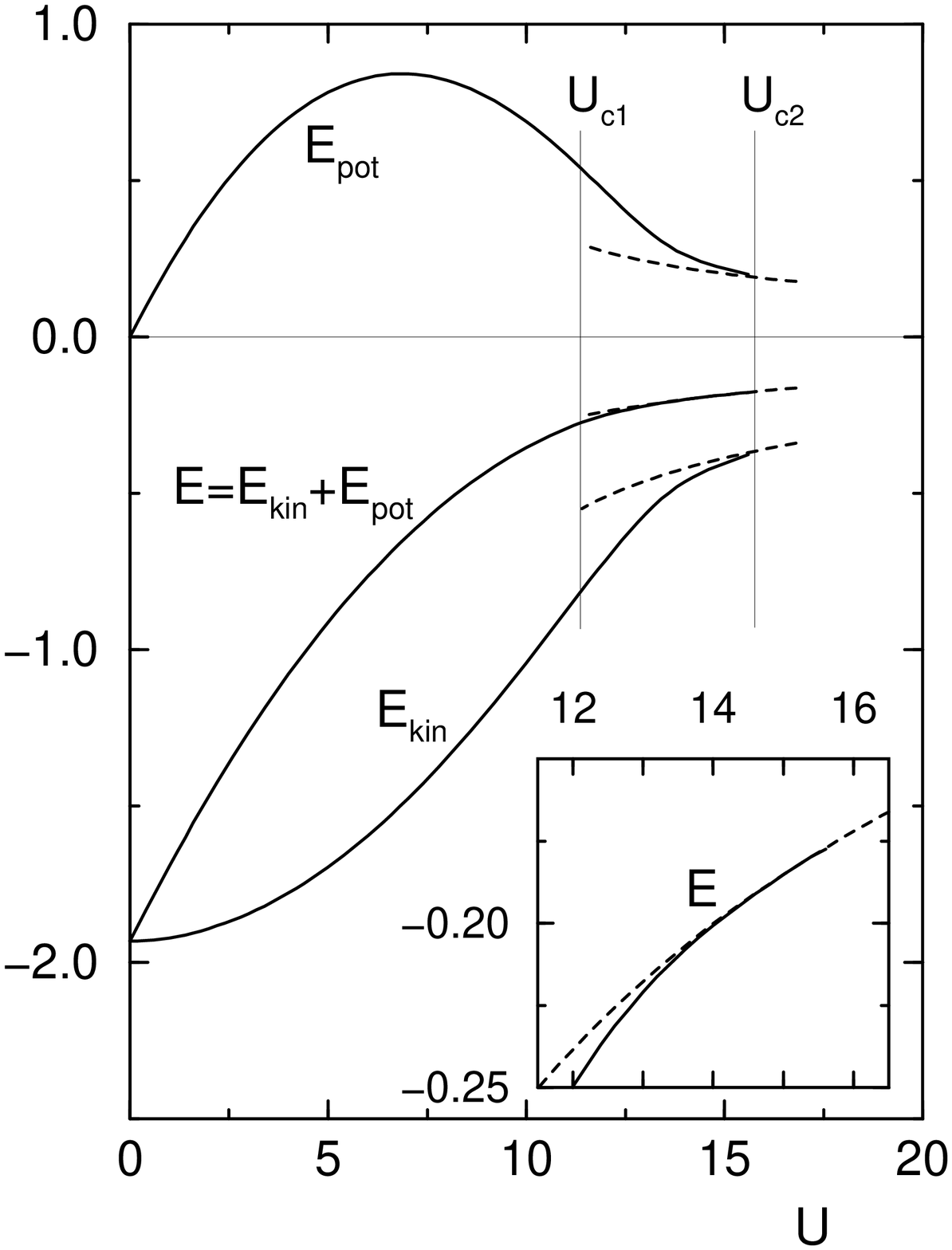,width=80mm,angle=0}}
\vspace{-3mm}

\parbox[]{85mm}{\small Fig.~10.
Kinetic, potential and total energy per site as functions of $U$ in 
the metallic (solid lines) and the insulating solution (dashed lines)
for the $d=5$ sc(100) film.
\label{fig:energy}
}
\end{figure}

\begin{figure}[t] 
\vspace{-3mm}
\center{\psfig{figure=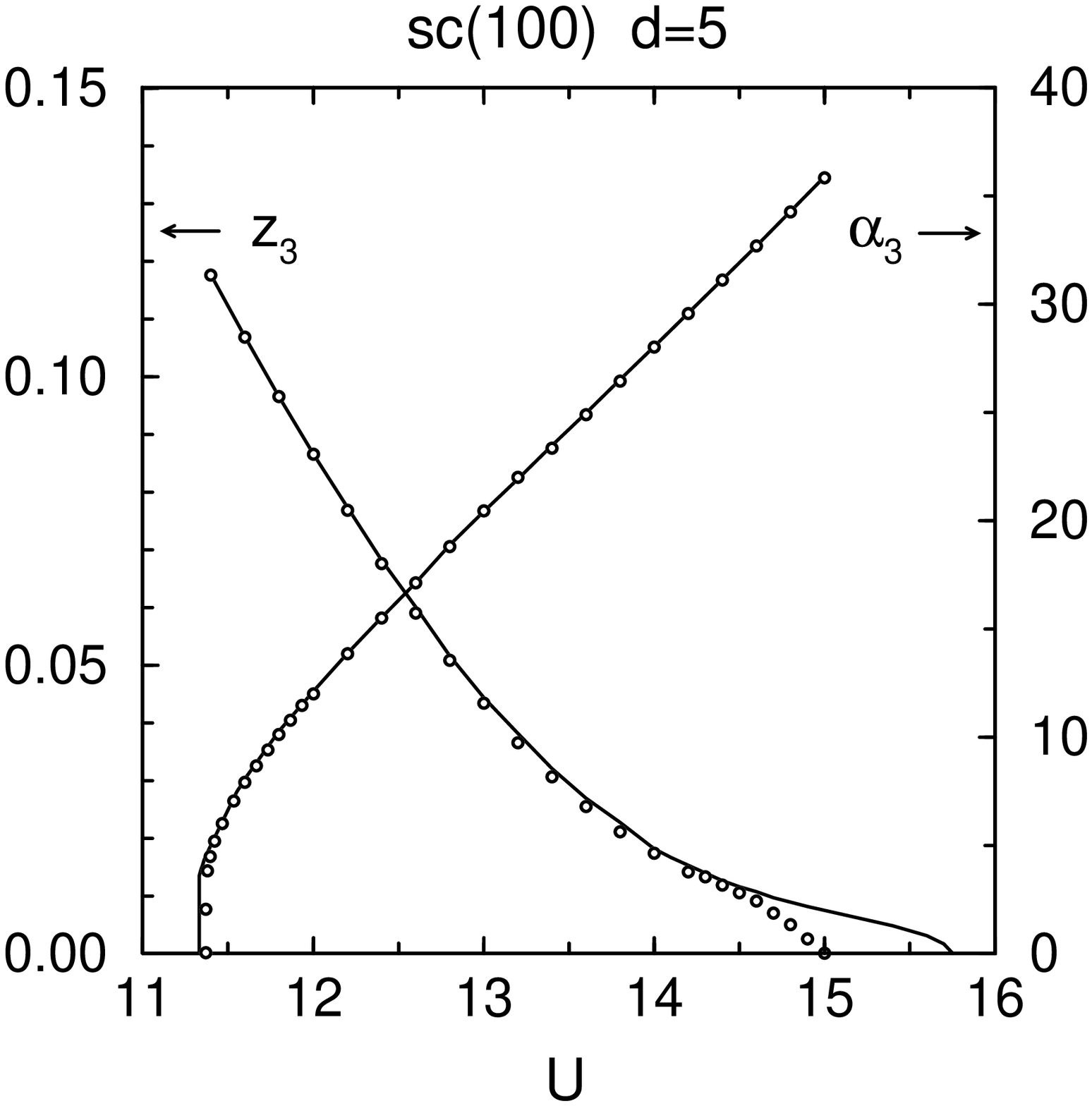,width=70mm,angle=0}}
\vspace{-3mm}

\parbox[]{85mm}{\small Fig.~11.
$U$ dependence of the quasi-particle weight $z_{3}$ and of 
the coefficient $\alpha_3$ in the central layer of the 
$d=5$ sc(100) film. Solid lines: $n_s=8$ (metal) and $n_s=7$
(insulator). Circles: $n_s=10$ (metal) and $n_s=9$
(insulator). 
\label{fig:conv}
}
\end{figure}

\begin{figure}[t] 
\vspace{-0mm}
\center{\psfig{figure=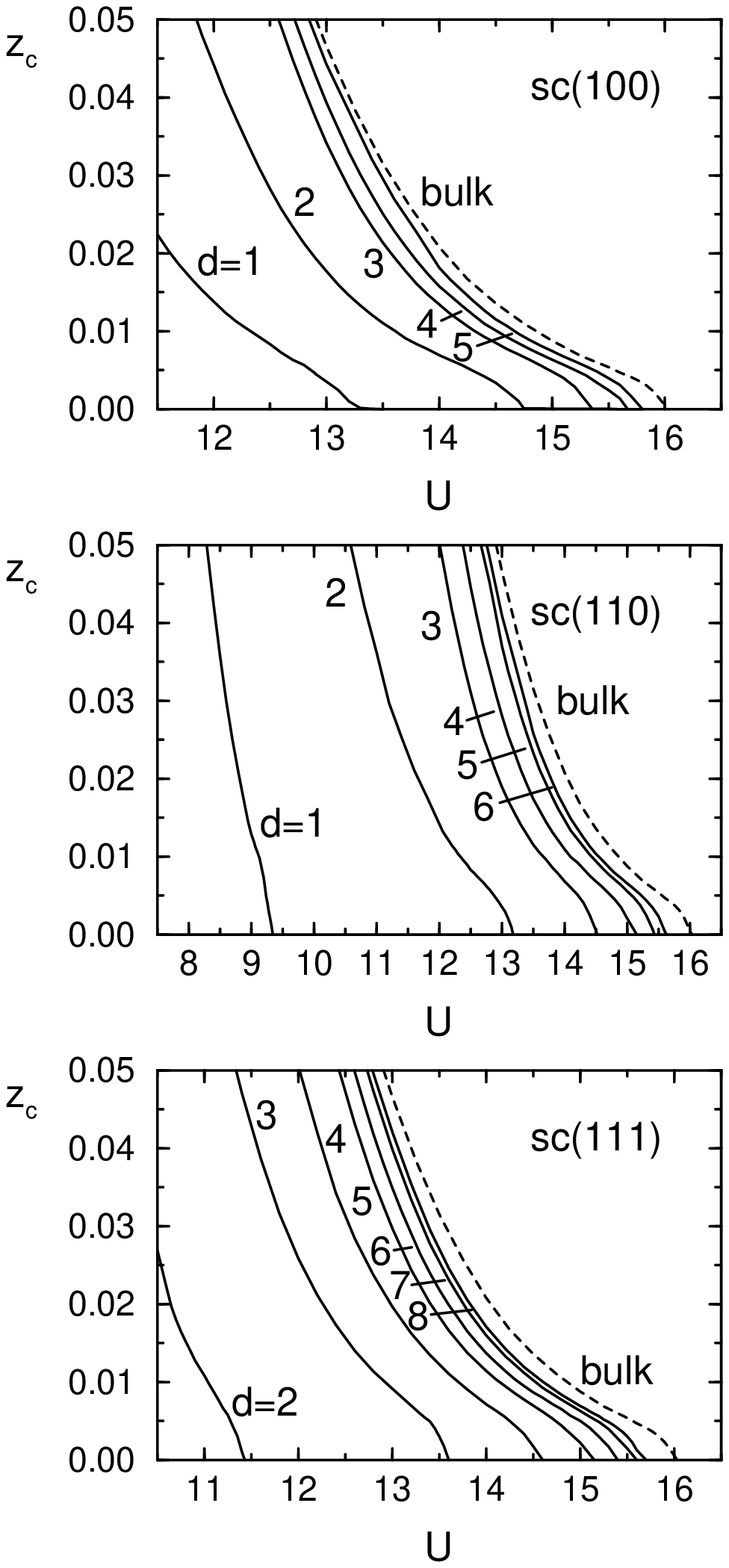,width=70mm,angle=0}}
\vspace{-0mm}

\parbox[]{85mm}{\small Fig.~12.
Quasi-particle weight of the central layer $\alpha=d/2$ 
($\alpha=(d+1)/2$, respectively) as a function of $U$ for 
sc(100), sc(110) and sc(111) 
films (solid lines) with a thickness $d$ ranging from $d=1$ 
(two-dimensional Hubbard model) up to $d=5$ ($d=6$, $d=8$,
respectively) 
and the result for $d\mapsto \infty$ (bulk, dashed line).
Note the different $U$ scales.
\label{fig:comb}
}
\end{figure}

The layer-dependent average double occupancy is shown in Fig.~9 as a 
function of $U$. For small $U$ and all layers, $\langle n_\uparrow
n_\downarrow \rangle$ decreases linearly as in the Brinkman-Rice
solution \cite{BR70}. At $U=U_{c2}$ it smoothly joins with the
(non-zero) average double occupancy of the insulating solution. For 
the insulator $\langle n_\uparrow n_\downarrow \rangle$ increases 
with decreasing $U$ down to $U_{c1}$. For both, the metal and the
insulator, double occupancies are suppressed more strongly at the 
film surfaces compared with the film center. Again, this is due to
the stronger effective Coulomb interaction $U/\sqrt{\Delta}$ which
results from the reduced variance $\Delta$ of the surface free
local density of states. 

Within the ED method a second-order transition from the metallic
to the insulating phase is predicted at $U=U_{c2}$. This can
be seen from the $U$ dependence of the respective internal energies:
A layer-dependent kinetic and potential energy per site
can be defined as:
\begin{eqnarray}
  T_\alpha &=& 
  \sum_{j\sigma} 
  t_{ij} \langle c^\dagger_{i\sigma} c_{j\sigma} \rangle
  \;\;\;\;\;\;\; (i\in \alpha)
  \nonumber \\
  V_\alpha &=&
  U \langle n_{i\uparrow} n_{i\downarrow} \rangle
  \;\;\;\;\;\;\; (i\in \alpha) \: .
\label{eq:e0}
\end{eqnarray}
The double occupancy $\langle n_{i\uparrow} n_{i\downarrow} \rangle$ 
and thus the
potential energy can be obtained from the (ED) solution of the 
impurity models directly. The kinetic energy may be calculated from:
\begin{equation}
  T_\alpha = \lim_{\beta \mapsto \infty} 
  \frac{4}{\beta} \mbox{Re} \sum_{n=0}^\infty
  (iE_n G(iE_n) - 1) + \frac{U}{2} 
  - 2 U \langle n_{i\uparrow} n_{i\downarrow} \rangle \: .
\label{eq:t0}
\end{equation}
Fig.~10 shows the kinetic
($E_{\rm kin}=(1/d) \sum_\alpha T_\alpha$),
potential 
($E_{\rm pot}=(1/d) \sum_\alpha V_\alpha$)
and total energy per site
($E=E_{\rm kin} + E_{\rm pot}$)
for the $d=5$ sc(100) film.
Due to the presence of the film surface, the absolute value of the 
kinetic energy per site 
at $U=0$ is somewhat lowered compared with kinetic energy per site
of the $D=3$ lattice which amounts to $E_{\rm kin}^{(D=3)} = -2.005$.
Comparing the different energies in the coexistence region
$U_{c1} < U < U_{c2}$, we notice that the gain in kinetic energy
is almost perfectly outweighted by the loss in potential energy
when passing from the insulating to the metallic solution at a
given $U$. A similar cancellation effect has been observed 
beforehand for the $D=\infty$ Bethe lattice \cite{RMK94}. 
Fig.~10 shows a remaining very small difference in the total energy
which favors the metallic solution in the entire coexistence region.

Although this is a consistent result within the ED method, it 
must be considered with some care: For the metallic solution 
close to $U_{c2}$ the relevant energy scale is determined by 
the width of the quasi-particle resonance near $E=0$. This width 
is approximately given by $z_\alpha W$ where $W$ is the free 
band width. Since $z_\alpha \mapsto 0$ for $U\mapsto U_{c2}$, 
one should expect that finite-size effects become important 
here and that an accurate determination of the internal energy 
cannot be achieved by the ED method. The same holds for the 
determination of $U_{c2}$ (and also of critical exponents). 
With the ED method we cannot access the very critical regime. 

Fig.~11 gives an example. Here we compare the quasi-particle 
weight $z_3$ at the center of the $d=5$ sc(100) film obtained 
for $n_s=8$ with the result for $n_s=10$. We notice that 
finite-size effects are unimportant for $z_s>0.01$. Generally, 
$n_s=8$ sites are sufficient for convergence if $U$ is well below 
$U_{c2}$. This can also be seen in Fig.~5 where a comparison with 
the results for $n_s=10$ is shown at a few points. For $z<0.01$, 
however, there are non-negligible differences: $U_{c2}$ is lowered 
by about $5\%$ when passing  from $n_s=8$ to $n_s=10$ (Fig.~11). 
This gives an estimate for the error in the determination of 
$U_{c2}$. The smallest reliable value for the quasi-particle 
weight $z_{\rm min}$ determines the ``energy resolution'' 
$\Delta E$ that can be achieved by the ED method for a given 
$n_s$. We have $\Delta E \approx z_{\rm min} W$. For $n_s=8$ 
and with $z_{\rm min}\approx 0.01$ this yields $\Delta E\approx 0.1$.
The error that is introduced by the finite low-energy cutoff is
of minor importance compared with the error due to finite-size 
effects. This is plausible since the smallest fictitious Matsubara 
frequency can be made as small as $\Delta E$ (here we have 
$\pi \widetilde{\beta}^{-1}\approx 0.05$). 

In the particle-hole symmetric case there is always one 
conduction-band energy with $\epsilon_k^{(\alpha)}=0$ (per layer). 
The corresponding hybridization strength $V_k^{(\alpha)}$ vanishes 
as $U \mapsto U_{c2}$ in the self-consistent calculation. 
Therefore, for the insulator at $U>U_{c2}$ we are effectively 
left with $n_s-1$ sites that are coupled via the hybridization
term. The metallic solution for $n_s=8$ ($n_s=10$) thus merges
with the insulating solution for $n_s=7$ ($n_s=9$) at $U_{c2}$.
Since the quasi-particle resonance is absent in the insulating
solution, there are no problems with the $n_s$ convergence in 
this case. The comparison between the results for $n_s=7$ and
$n_s=9$ is also shown in Fig.~11. The critical interaction $U_{c1}$
can be determined much more precisely.

\begin{figure}[t] 
\vspace{-0mm}
\centerline{\psfig{figure=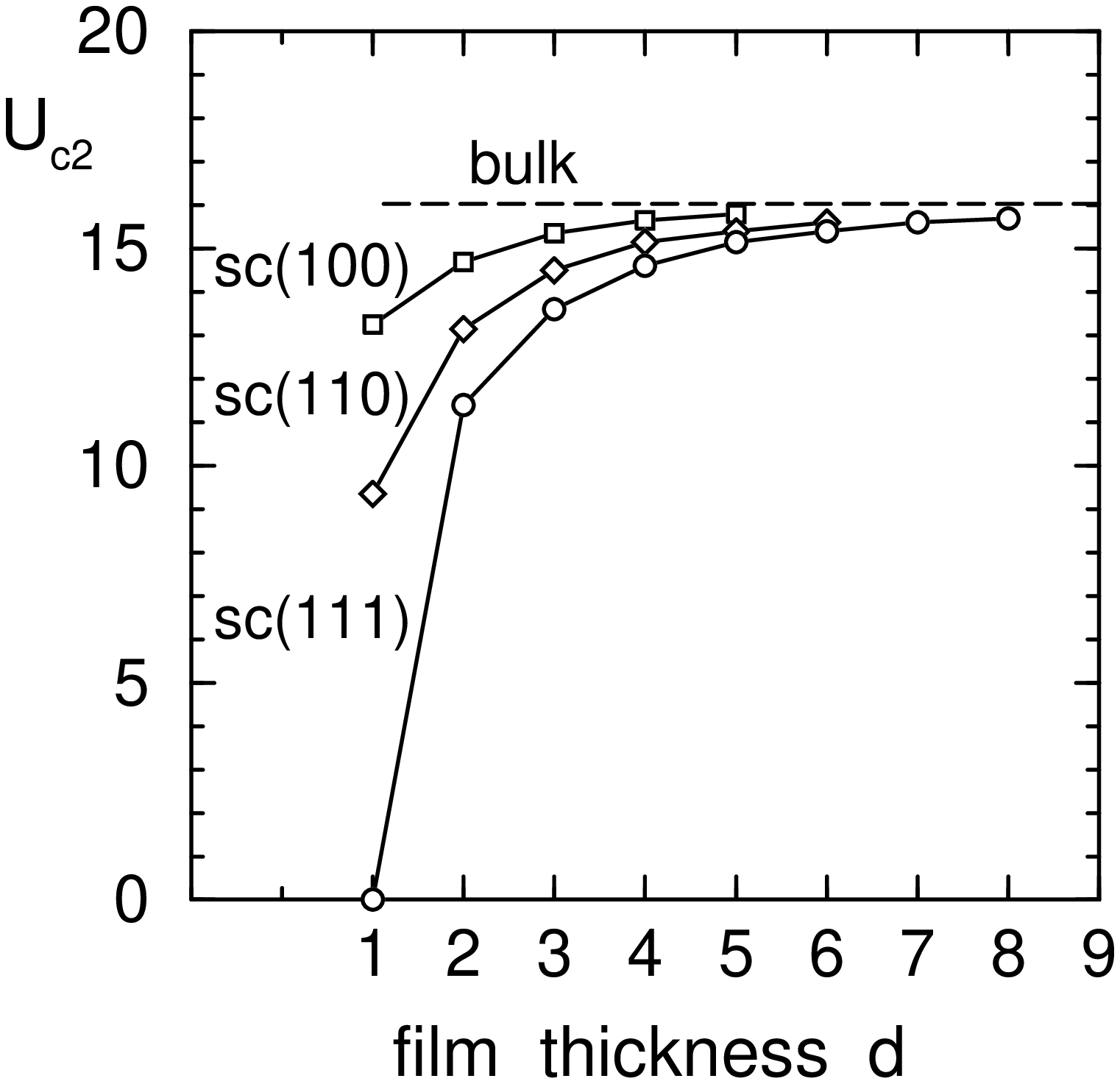,width=80mm,angle=0}}
\vspace{0mm}

\parbox[]{85mm}{\small Fig.~13.
Thickness dependence of the critical interaction $U_{c2}$ 
for sc(100), sc(110) and sc(111) films.
\label{fig:uc2}
}
\end{figure}

\begin{figure}[t] 
\vspace{-0mm}
\center{\psfig{figure=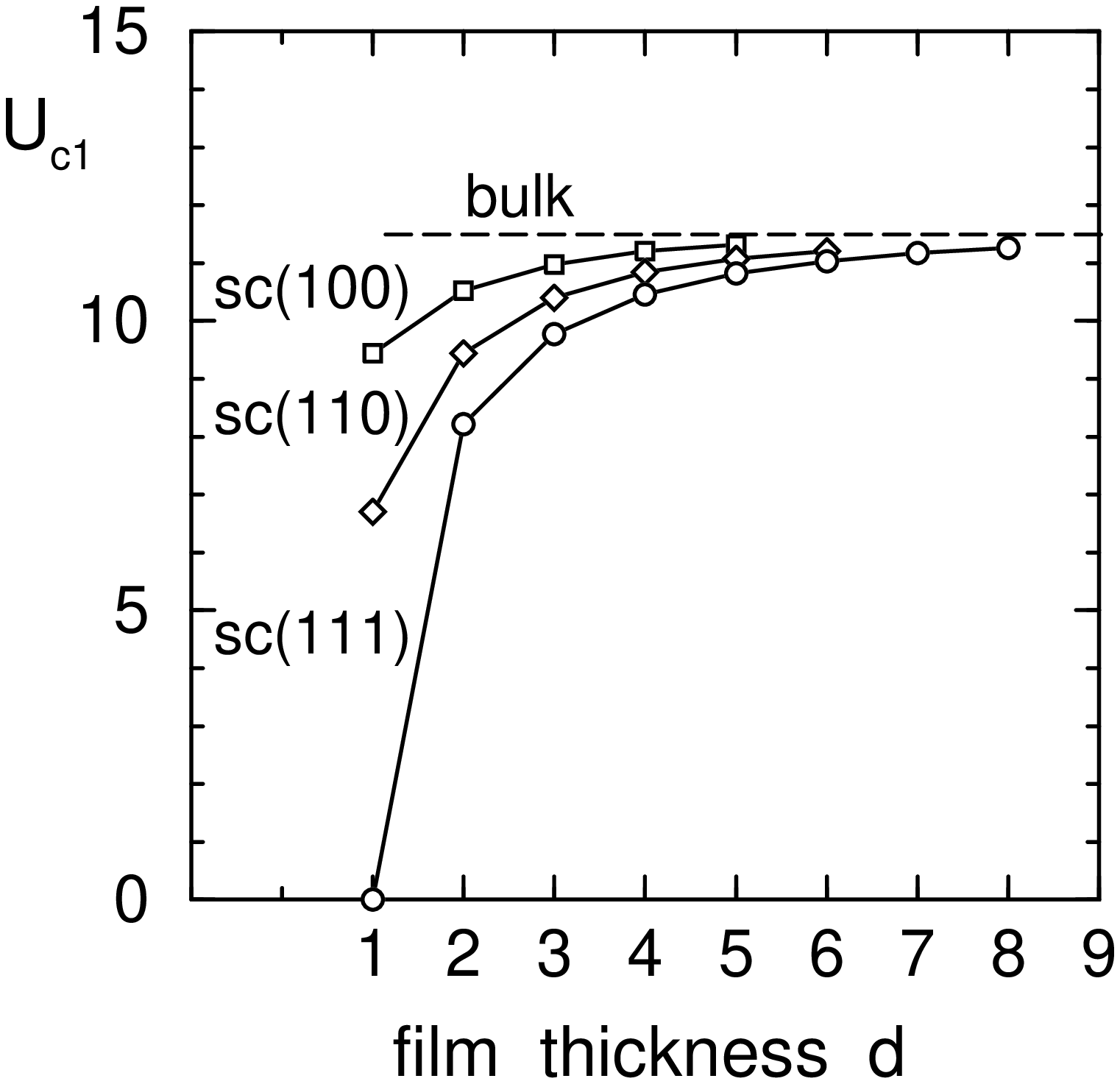,width=80mm,angle=0}}
\vspace{0mm}

\parbox[]{85mm}{\small Fig.~14.
Thickness dependence of $U_{c1}$.
\label{fig:uc1}
}
\end{figure}

A basic question in a study of the Mott transition in thin films
concerns the thickness dependence of the critical interactions 
$U_{c2}$ and $U_{c1}$. Since it is the general trend that is of
primary interest, the mentioned ambiguity in the determination of
$U_{c2}$ plays a minor role only. For all geometries and 
thicknesses considered, we found $U_{c2}$ and $U_{c1}$ to be
unique, i.~e.\ for a given system all layer-dependent quasi-particle
weights and also all $1/E$ coefficients vanish at a common
critical value of the interaction, respectively. It is thus
sufficient to concentrate on the quantities at the film center.

Fig.~12 shows the central-layer quasi-particle weight $z_c$ as 
a function of $U$ for the different systems in the region
close to $U_{c2}(d)$. It is worth mentioning that for all 
three film geometries and for all $d\ge 2$ the functions $z_c(U)$ 
are more or less only shifted rigidly to lower interactions
compared with the result $z(U)$ for the bulk (which of course
is independent from the choice of the film surface). This is
in clear contrast to the trend of $z_c(U)$ for weaker interactions
(see the result for the sc(111) film in Fig.~5, for example):
In the critical regime but also at weaker interactions where 
finite-size effects are unimportant, the $U$ dependence of 
$z_\alpha$ is rather insensitive to details of the film geometry. 
On the other
hand, the shift of $z_c(U)$ with respect to $z(U)$ does depend
on the thickness $d$. In all cases $z_c(U)$ shifts to higher
interaction strengths with increasing $d$ and converges to the 
bulk curve $z(U)$ finally. This also implies that without any 
exception $z_c$ increases with increasing $d$ at a given $U$. 
We can state that the general trends are remarkably simple.

From the zero of $z_c(U)$ we can determine the critical interaction
$U_{c2}$ for a given geometry and thickness. The results for the
different systems are summarized in Fig.~13. $U_{c2}$ is a 
monotonously increasing function of $d$ for the sc(100), the 
sc(110) as well as for the sc(111) films. For a finite thickness, 
$U_{c2}(d)$ is always smaller than the bulk value
which is apparently approached only in the limit $d\mapsto \infty$.
For a given film thickness, $U_{c2}$ increases when passing from
the sc(111) to the sc(110) and to the sc(100) film. (Note that 
for the $d=1$ sc(111) film $U_{c2}=0$ since there is no 
intra-layer hopping in this case.) The convergence with respect
to $d$ is the fastest for the sc(100) and the slowest for the
sc(111) films. These trends seem to be related to the differences
in the film-surface coordination numbers: $n_{(100)}=5$, 
$n_{(110)}=4$, $n_{(111)}=3$.

The interaction strength at which the central-layer $1/E$
coefficient $\alpha_c$ drops to zero determines $U_{c1}$ (see
Fig.~7). Its thickness dependence is shown in Fig.~14 for the 
different geometries. The result is rather surprising: Apart
from small uncertainties in the determination of $U_{c2}$ and
$U_{c1}$, the relative thickness and geometry dependence of
$U_{c1}$ is the same as that of $U_{c2}$. Only the 
absolute values for $U_{c1}$ are smaller: $U_{c1}(d)$ converges
to the bulk value $U_{c1}=11.5$ which is well below $U_{c2}=16.0$.
If we rescale $U_{c1}(d)$ for the different geometries by the same 
(bulk) factor $r=U_{c2}/U_{c1} \approx 1.39$, we end up with the 
results for $U_{c2}(d)$ shown in Fig.~13 within a tolerance
($|r \cdot U_{c1}(d) - U_{c2}(d)|/ U_{c2} < 0.005$) that is much
smaller than e.~g.\ the error in the determination of $U_{c2}$ 
due to finite-size effects. The latter is irrelevant here if one 
assumes $U_{c2}$ to be overestimated by the same constant factor, 
for the bulk and for the films, which would only change the ratio 
$r$. Looking at the trends in the results for $n_s=8$ in Fig.~12, 
this assumption is quite plausible. The found relation between 
$U_{c2}$ and $U_{c1}$ is surprising because the disappearance of 
the insulating solution for $U\mapsto U_{c1}$ is expected to be of 
different nature compared with the breakdown of the Fermi-liquid 
metallic phase as $U\mapsto U_{c2}$. 

For $U_{c2}$, all the details of its geometry and thickness 
dependence can be understood by a simple but instructive argument. 
The main idea has first been developed by Bulla \cite{bulla} for 
the $D=\infty$ case. Here we discuss a generalization for the film 
geometries: 

The argument assumes that for $U\mapsto U_{c2}$ one can 
disregard the effects of the high-energy charge excitation peaks at 
$E\approx \pm U/2$ and that the quasi-particle resonance near $E=0$ 
results from a SIAM hybridization function $\Delta^{(\alpha)}(E)
= \sum_k (V_k^{(\alpha)})^2/(E-\epsilon_k^{(\alpha)})$ that consists
of a single pole at $E=0$:
\begin{equation}
  \Delta^{(\alpha)}(E) \mapsto \frac{\Delta_N^{(\alpha)}}{E} \: .
\label{eq:delta}
\end{equation}
With the index $N$ we refer to the $N$-th step in the iterative 
solution of the DMFT self-consistency equation. For each layer 
$\alpha=1,...,d$ the one-pole structure of $\Delta^{(\alpha)}(E)$
corresponds to an $n_s=2$ site SIAM with a conduction-band energy
at $\epsilon^{(\alpha)} = U/2$ and hybridization strength
$V^{(\alpha)} = (\Delta_N^{(\alpha)})^{1/2}$. The analytic
solution up to second order in $V^{(\alpha)}/U$ 
(cf.\ Ref.\ \cite{Hew93})
yields two peaks in the impurity spectral function at 
$E \approx \pm U/2$ as well as two peaks near $E=0$ which build up 
the Kondo resonance for $U\mapsto U_{c2}$. The weight of the
resonance is thus given by:
\begin{equation}
  z_\alpha = 2\cdot \frac{18 (V^{(\alpha)})^2}{U^2} 
  = \frac{36}{U^2} \Delta_N^{(\alpha)} \: .
\label{eq:qp}
\end{equation}
In the self-consistent solution this is also the layer-dependent 
quasi-particle weight which determines the coherent part of the 
film Green function in the low-energy regime via Eq.\ (\ref{eq:gk}). 
The coherent part of the on-site Green function in the $\alpha$-th 
layer may be written as:
\begin{equation}
  G_{\alpha}^{\rm (coh.)}(E) = z_\alpha \cdot
  \widetilde{G}_\alpha(E) = 
  \frac{z_\alpha}{E-\widetilde{M}_\alpha^{(2)} F_\alpha(E)} \: .
\label{eq:cf1}
\end{equation}
Here, $\widetilde{G}_\alpha(E)$ is the on-site element of the free
film Green function, but calculated for the renormalized hopping 
matrix $\epsilon_{\alpha \beta}({\bf k}) \mapsto \sqrt{z_\alpha} \,
\epsilon_{\alpha \beta}({\bf k}) \sqrt{z_\beta}$ (see Eq.\ 
(\ref{eq:gk})). The second expression represents the first step 
in a continued-fraction expansion which involves the second moment
$\widetilde{M}_\alpha^{(2)} = \int \,dE \, E^2 \, 
\widetilde{G}_\alpha(E)$ of $\widetilde{G}_\alpha(E)$. For the 
remainder we have $F_\alpha(E) = 1/E + {\cal O}(E^{-2})$. 

Starting from Eq.\ (\ref{eq:delta}) in the $N$-th step, the DMFT 
self-consistency equation (\ref{eq:dmft}) yields a new hybridization 
function via $\Delta^{(\alpha)}(E)=E-(\epsilon_d-\mu)-\Sigma_\alpha(E)
- G_\alpha(E)^{-1}$. Using $\epsilon_d = 0$, $\mu = U/2$, 
$\Sigma_\alpha(E) = U/2 + (1-z_\alpha^{-1}) E + \cdots$ and
Eq.\ (\ref{eq:cf1}), we get:
\begin{equation}
  \Delta^{(\alpha)}(E) =  \frac{1}{z_\alpha} 
  \widetilde{M}_\alpha^{(2)} F_\alpha(E) 
\label{eq:deltacf}
\end{equation}
for energies close to $E=0$. Insisting on the one-pole structure
of $\Delta^{(\alpha)}(E) \stackrel{!}{=} \Delta^{(\alpha)}_{N+1}/E$ 
for $U\mapsto U_{c2}$, we must have $F_\alpha(E) = 1/E$. This 
amounts to replacing the coherent part of the film Green function
by the simplest Green function with the same moments up to the 
second one. To express
$\Delta^{(\alpha)}_{N+1}$ in terms of $\Delta^{(\alpha)}_{N}$, we 
still need $\widetilde{M}_\alpha^{(2)}$. Let us introduce the 
intra- and inter-layer coordination numbers $q$ and $p$ (e.g.\ 
$q=4$ and $p=1$ for the sc(100) geometry). The second moment is 
easily calculated by evaluating an (anti-)commutator of the form
$\langle [ \,[[c,\widetilde{H}_0]_-,\widetilde{H}_0]_-,c^\dagger]_+
\rangle$. This yields:
\begin{equation}
  \widetilde{M}_\alpha^{(2)} = z_\alpha 
  (q z_\alpha + p z_{\alpha-1} + p z_{\alpha+1}) \, t^2 \: .
\label{eq:moment}
\end{equation}
We also define the following tridiagonal matrix with dimension $d$:
\begin{equation}
  {\bf K}(U) = \frac{36 \, t^2}{U^2} \left( 
  \begin{array}{ccccc}
  q & p &        &        &\\
  p & q & p      &        &\\
    & p & q      & .. &\\
    &   & .. & .. &\\
  \end{array} \right) \: .
\label{eq:kmatrix}
\end{equation}
Inserting (\ref{eq:moment}) and (\ref{eq:qp}) into 
(\ref{eq:deltacf}), yields a ``linearized'' self-consistency 
equation for $U\mapsto U_{c2}$:
\begin{equation}
  \Delta^{(\alpha)}_{N+1} = \sum_\beta K_{\alpha \beta}(U) \:
  \Delta^{(\beta)}_{N} \: .
\label{eq:lindmft}
\end{equation}
A fixed point of ${\bf K}(U)$ corresponds to a 
self-consistent solution. Let $\lambda_r(U)$ denote the 
eigenvalues of ${\bf K}(U)$. We can distinguish between two 
cases: If $|\lambda_r(U)| < 1$ for all $r=1,...,d$, there is the
trivial solution $\lim_{N\mapsto \infty} \Delta^{(\alpha)}_{N}
= 0$ only. This situation corresponds to the insulating solution
for $U>U_{c2}$. Contrary, if there is at least one $\lambda_r(U)>1$,
$\Delta^{(\alpha)}_{N}$ diverges exponentially as $N\mapsto \infty$.
This indicates the breakdown of the one-pole model for the 
hybridization function in the metallic solution for $U<U_{c2}$. 
The critical interaction is thus determined by the maximum 
eigenvalue:
\begin{equation}
  \lambda_{\rm max}(U_{c2}) = 1 \: .
\label{eq:cond}
\end{equation}
The eigenvalues of a tridiagonal matrix (\ref{eq:kmatrix}) can be 
calculated analytically for arbitrary matrix dimension $d$
\cite{Ste90}: The $\lambda_r(U)$ are the zeros of the $d$-th
degree Chebyshev polynomial of the second kind. Solving 
(\ref{eq:cond}) for $U_{c2}$ then yields:
\begin{equation}
  U_{c2}(d,q,p) 
  = 6 t \: \sqrt{q+2p \cos\left(\frac{\pi}{d+1}\right)} \: .
\label{eq:uc}
\end{equation}

\begin{figure}[t] 
\vspace{-0mm}
\center{\psfig{figure=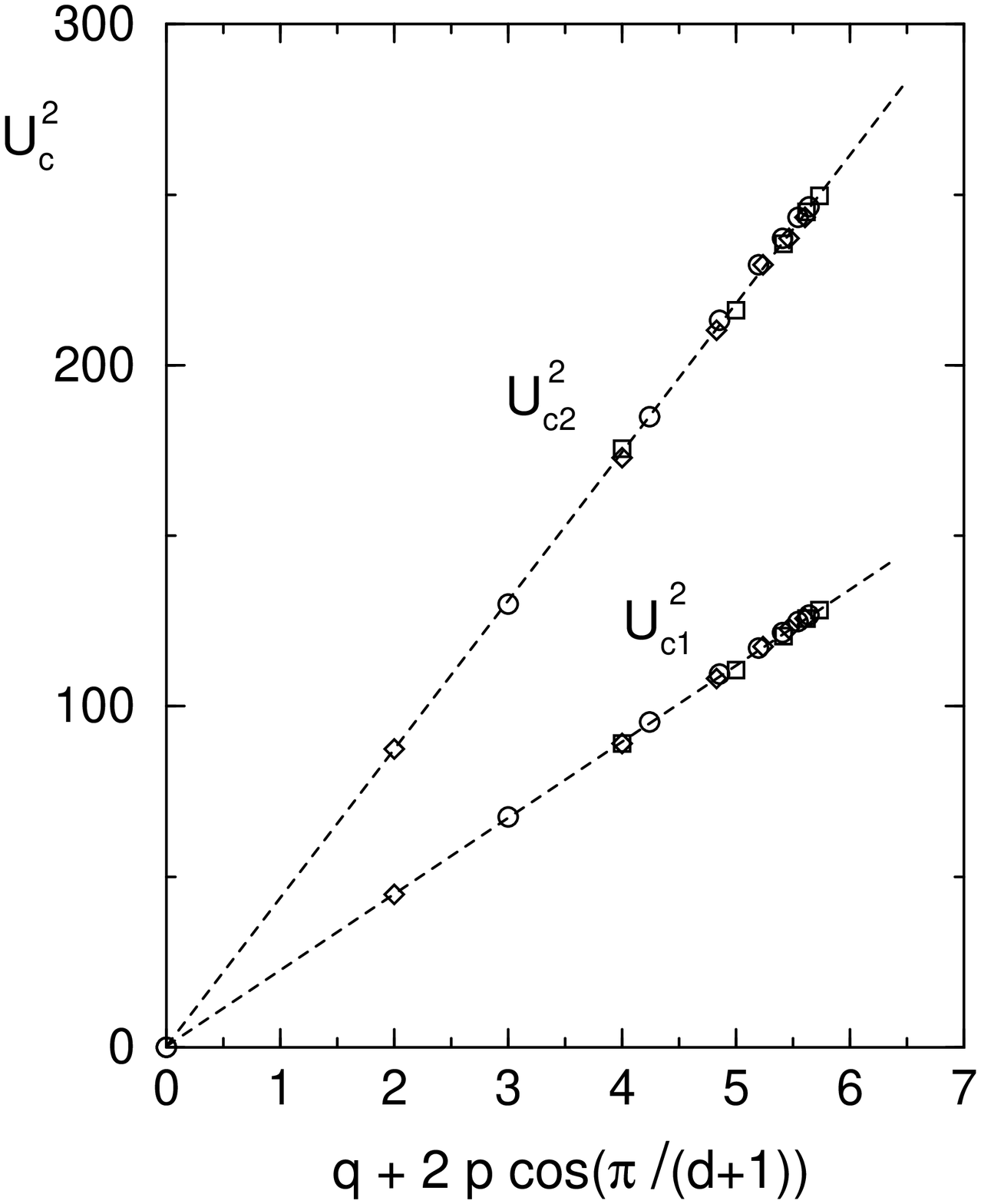,width=70mm,angle=0}}
\vspace{4mm}

\parbox[]{85mm}{\small Fig.~15.
Critical interactions $U^2_{c2}$ and $U^2_{c1}$ 
for sc(100) (squares), sc(110) (diamonds) 
and sc(111) films (circles) as in Fig.~13 and Fig.~14, 
but plotted against $q+2p\cos(\pi/(d+1))$. The dashed lines
are linear fits to the data.
\label{fig:rg}
}
\end{figure}

Fig.~15 shows that the numerical results for $U_{c2}$ can be
well described by this simple formula. Plotting $U_{c2}^2$
against $q+2p\cos(\pi/(d+1))$, yields a linear dependence as 
a good approximation. The slope in the linear fit to the data,
however, turns out to be $(6.61 t)^2 > (6t)^2$. This shows that
there is a {\em systematic} overestimation of the critical
interaction by about 10\% compared with Eq.\ (\ref{eq:uc}). 
Partially, this has to be ascribed to finite-size effects
in the ED method since the error estimated from the results in
Fig.~11 is of the same order of magnitude. (As mentioned before,
the fact that the error is a systematic one, is explained by
the very simple trends seen in Fig.~12). On the other hand, 
we also have to bear in mind that (\ref{eq:uc}) rests on some
simplifying assumptions.

While we have achieved a satisfactory understanding of the 
results for $U_{c2}$, it remains unclear to us why
Eq.\ (\ref{eq:uc}) (with the factor $6t$ replaced by a suitable
constant) also well describes the thickness and geometry
dependence of $U_{c1}$. Again, Fig.~15 shows a linear trend.
The slope is $(4.74 t)^2$. The above-developed argument, however,
obviously breaks down.

It is instructive to compare these results with the analogous results
for a thermodynamic phase transition in thin films. Within the
mean-field approach, the layer-dependent magnetization $m_\alpha$
in Ising films with coupling constant $J$ is determined by the 
self-consistency equation:
\begin{equation}
  m_\alpha = \tanh \left(
  \frac{J}{2 k_B T} (q m_\alpha + p m_{\alpha+1} + p m_{\alpha-1})
  \right) \: .
\end{equation}
The equation can be linearized for temperatures $T$ near the 
Curie temperature $T_C$ where $m_\alpha \ll 1$. Comparing with
Eq.\ (\ref{eq:lindmft}) yields the following analogies: 
\begin{equation}
m_\alpha \Leftrightarrow z_\alpha
\: , \;\;\;
J/2 \Leftrightarrow 36t^2
\: , \;\;\;
k_B T \Leftrightarrow U^2
\: , \;\;\;
k_B T_C \Leftrightarrow U_{c2}^2
\: .
\end{equation}
The exact Curie temperature in thick ($d\mapsto \infty$) Ising 
films is expected \cite{FF67,All70} to obey the power law 
$(T_c(\infty) - T_c(d))/T_c(\infty) = C_0d^{-\lambda}$ where 
the shift exponent $\lambda=1/\nu$ is related to the ($D=3$) 
critical exponent $\nu$ of the correlation function. Within the 
mean-field approach to the magnetic phase transition in Ising
films one obtains $\lambda = 2$.

The same exponent is found within dynamical mean-field theory
applied to the Mott transition in Hubbard films. Expanding 
Eq.\ (\ref{eq:uc}) in powers of $1/d$, we obtain:
\begin{equation}
  \frac{U_{c2} - U_{c2}(d,q,p)}{U_{c2}} = 
  C_0(q,p) \cdot d^{-\lambda} \: ,
\end{equation}
where $U_{c2}$ denotes the bulk value, $C_0(q,p) = \pi^2 q/(q+2p)$ 
and $\lambda = 2$ the exponent.
\\

{\center \bf \noindent IV. CONCLUSION \\ \mbox{} \\} 

By the invention of dynamical mean-field theory we are in a 
position to treat itinerant-electron models on the same footing 
as does the Weiss molecular-field theory with respect to 
localized spin models. To gain a first insight into collective 
magnetism for thin-film and surface geometries, the molecular-field 
approach has successfully been applied to the Ising or the Heisenberg 
model in the past. It is well known that the reduced system symmetry 
may result in characteristic modifications of the magnetic properties 
which are interesting on their own but also with respect to technical 
applications. The Weiss theory is able to describe a major number
of magnetic properties qualitatively correct.

To study the correlation-induced Mott transition from a paramagnetic
metal to a paramagnetic insulator, we need to consider an 
itinerant-electron model. Presumably, the Hubbard model is
the simplest one in this respect. Just as the Weiss theory 
for the magnetic properties of Ising and Heisenberg films, the
DMFT provides the first step in an understanding of the Mott
transition and the related electronic properties in Hubbard films.
This has been the central idea of the present study. We have
generalized the mean-field equations for the application to systems
with reduced translational symmetry and have solved them using
the exact-diagonalization method. Let us briefly list up the main 
results found:

Similar to the results for the infinite-dimensional Bethe lattice, 
we find a metallic and an insulating solution of the mean-field
equations at $T=0$ for all film geometries considered. These coexist 
in a certain range of the interaction strength. The metallic solution 
is stable against the insulating one in the whole coexistence region 
(however, in the critical regime the ED approach is questionable).

Generally, the breakdown of translational symmetry with respect to 
the film normal direction leads to modifications of the electronic 
structure. For film thickness $d \mapsto \infty$ these disappear 
in the bulk, while surface effects are still present. The finite film
thickness as well as the presence of the surface manifest themselves 
in a significant layer dependence of the on-site Green function and
the self-energy for both, the metallic and the insulating solution.
The layer dependence is found to be the more pronounced the more 
open is the film surface. In thicker films surface effects quickly 
diminish when passing from the top layer to the film center.

In particular, we have considered the so-called layer-dependent 
quasi-particle weight $z_\alpha$ for the metallic phase as a function 
of $U$, $d$ and geometry. For $U\ne 0$, $z_\alpha<1$ is the reduction 
factor of the discontinuous drop of the momentum-distribution function
in the $\alpha$-th layer at each of the $d$ one-dimensional 
Fermi-``surfaces'' or, equivalently, the weight of the coherent 
quasi-particle peak in the local density of states of the $\alpha$-th
layer. In all cases the quasi-particle weight at the film surfaces
is found to be significantly reduced which is due to the surface
enhancement of the effective Coulomb interaction $U/\sqrt{\Delta}$ 
($\Delta$ is the variance of the free local density of states). 
At the film surfaces the electrons are ``heavier'', double occupancies 
are suppressed more effectively. The layer dependence of the 
quasi-particle weight is oscillatory in some cases at small and 
intermediate interaction strengths. As $U$ approaches the critical 
regime, the behavior changes qualitatively;
here $z_\alpha$ is always monotonously increasing with increasing
distance from a film surface. Furthermore, for fixed $U$, $z_\alpha$
always increases with increasing $d$ -- the general trends are
remarkably simple.

The low-energy electronic structure in the insulating solution is
governed by the $1/E$ coefficient in the low-energy expansion of 
the self-energy. Generally, the layer dependence of the coefficient
is less spectacular compared with the layer dependence of $z_\alpha$ 
in the metallic solution. At the film surfaces the $1/E$ coefficient 
is somewhat enhanced and in all cases monotonously decreases with
increasing distance from a surface. 

For a given film geometry and thickness there is a unique critical 
interaction $U_{c2}$ at which all (layer-dependent) effective 
masses $z_\alpha^{-1}$ diverge. This implies that the whole film
is either metallic or a Mott insulator. For all cases investigated,
we did not find a surface phase that differs from the bulk phase.
This may change if non-uniform model parameters, e.~g.\ a modified 
surface $U$, are considered. The question is left for future
investigations. 

While a precise determination of the critical interaction is not
possible by means of the ED approach, general trends can be derived
safely. It is found that $U_{c2}$ is strongly geometry and 
thickness dependent. It monotonously increases with increasing
film thickness and smoothly approaches the bulk value from below. 
For finite $d$, $U_{c2}$ is the smaller and the convergence to the 
bulk value is the slower the more open is the film surface. The 
same trends are also seen for the critical interaction $U_{c1}$ at
which the insulating solution disappears. In fact, within numerical
accuracy a simple relation, $U_{c2}(d) = r \cdot U_{c1}(d)$, seems 
to hold. The geometry and thickness dependence of $U_{c2}$ can be 
understood qualitatively by (approximately) linearizing the 
mean-field equations for $U \mapsto U_{c2}$. After proper rescaling,
the resulting analytical expression for $U_{c2}(d,q,p)$ can even 
quantitatively reproduce the numerical data. The argument, however,
is not applicable to the critical interaction $U_{c1}$ and thus
cannot explain the ``empirically'' found relation between $U_{c1}$
and $U_{c2}$. An effective, linearized theory that is valid for 
$U \mapsto U_{c1}$ remains to be constructed. Calculating the Curie 
temperature of Ising films within the molecular-field theory, 
yields exactly the same trends. This analogy suggests that the 
observed geometry and thickness dependence may be {\em typical} 
for the mean-field treatment of the problem. Whether or not there 
are qualitative changes if $U_{c1,2}(d,q,p)$ could be calculated 
beyond mean-field theory, remains to be another open problem.

Finally, one could think of an experimental investigation of the 
Mott transition in thin films. The present study shows that the 
strong thickness dependence of the critical interaction has its 
origin in the reduced coordination number at the film surface.
A monotonous increase of the critical interaction with increasing 
$d$ should thus be expected also for temperatures above the N\'eel
temperature where we have a magnetically disordered
state. Studying the Mott transition in a bulk material, one needs 
to ``vary $U$'' (or temperature) experimentally. Contrary, for a 
thin-film geometry the transition may take place also by varying 
$d$. If it is possible to grow crystalline films of a metallic 
material which in the bulk is close to the Mott transition, one 
may observe insulating behavior in ultrathin films and a transition 
to a metallic phase with increasing thickness.
\\

{\center \bf \noindent Acknowledgements \\ \mbox{} \\} 

The authors would like to thank R.~Bulla (MPI f\"ur Physik
komplexer Systeme, Dresden) for helpful discussions and for
drawing our attention to Ref.\ \cite{bulla} prior to publication.
This work is supported by the Deutsche Forschungsgemeinschaft 
within the SFB 290.
\vspace{5mm}

---------------------------------------------------------------------

\end{document}